\documentstyle[prd,aps,eqsecnum,preprint,tighten]{revtex}

\input epsf

\begin{document}
\preprint{\vbox to 108 pt
{\hbox{IHES/P/96/13}\hbox{IASSNS-HEP 96/14}\hbox{BRX
TH-391}\hbox{CPT-95/P.E.3279}\hbox{gr-qc/9602056}\vfil}}
\draft
\title{Tensor--scalar gravity and binary-pulsar experiments}
\author{Thibault Damour\cite{DARC}}
\address{Institut des Hautes Etudes Scientifiques, F 91440
Bures-sur-Yvette, France,\\
and School of Natural Sciences, Institute for Advanced Study,\\
Olden Lane, Princeton, NJ 08540, USA}
\author{Gilles Esposito-Far\`ese\cite{CPT}}
\address{Department of Physics, Brandeis University, Waltham, MA
02254, USA}
\date{February 26, 1996}
\maketitle
\begin{abstract}
Some recently discovered nonperturbative strong-field effects in
tensor--scalar theories of gravitation are interpreted as a scalar
analog of ferromagnetism: ``spontaneous scalarization''. This
phenomenon leads to very significant deviations from general
relativity in conditions involving strong gravitational fields,
notably binary-pulsar experiments. Contrary to solar-system
experiments, these deviations do not necessarily vanish when the
weak-field scalar coupling tends to zero. We compute the scalar
``form factors'' measuring these deviations, and notably a parameter
entering the pulsar timing observable $\gamma$ through
scalar-field-induced variations of the inertia moment of the pulsar.
An exploratory investigation of the confrontation between
tensor--scalar theories and binary-pulsar experiments shows that
nonperturbative scalar field effects are already very tightly
constrained by published data on three binary-pulsar systems. We
contrast the probing power of pulsar experiments with that of
solar-system ones by plotting the regions they exclude in a generic
two-dimensional plane of tensor--scalar theories.
\end{abstract}
\narrowtext
\section{Introduction}
Einstein's general relativity theory postulates that gravity is
mediated only by a long-range tensor field. It has been repeatedly
pointed out over the years (starting with Kaluza \cite{K21}) that
unified theories naturally give rise to long-range scalar fields
coupled to matter with gravitational strength. This led many authors,
notably Jordan \cite{J+}, Fierz \cite{F56}, and Brans and Dicke
\cite{BD61}, to study, as most natural alternatives to general
relativity, tensor--scalar theories in which gravity is mediated in
part by a long-range scalar field. The motivation for such theories
has been recently revived by string theory which contains massless
scalars in its gravitational sector (notably the model-independent
dilaton).

We shall consider tensor--scalar gravitation theories containing only
one scalar field, assumed to couple to the trace of the
energy-momentum tensor. The simplest example of such a theory is a
scalar field only coupled to the gravitational sector through a
nonminimal coupling $\xi R \Phi^2$ (see Section \ref{SEC6} below).
For a study of the observable consequences of general tensor--scalar
theories (containing one or several scalar fields), see Ref.
\cite{DEF1}.

Actually, one generically expects scalar fields not to couple exactly
to the mass but to exhibit some ``composition dependence'' in their
couplings to matter. However, a recent study of a large class of
viable string-inspired tensor--scalar models \cite{DP94} has found
that the composition-dependent effects represent only a very small
fraction ($\sim 10^{-5}$) of the effective coupling to matter. Such
fractionally small composition-dependent effects would be negligible
in the gravitational physics of neutron stars that we consider here.

The most general theory describing a mass-coupled long-range scalar
contains one arbitrary ``coupling function'' $A(\varphi)$ \cite{F56}.
The action defining the theory reads
\begin{eqnarray}
S & = & {c^4\over 16 \pi G_*}\int {d^4 x\over c} g_*^{1/2}\left(
R_* - 2 g^{\mu\nu}_* \partial_\mu\varphi \partial_\nu\varphi\right)
\nonumber \\
&& +S_m[\psi_m; A^2(\varphi)g^*_{\mu\nu}]\ .
\label{eq1.1}\end{eqnarray}
Here, $G_*$ denotes a bare gravitational coupling constant,
$R_*\equiv g_*^{\mu\nu} R^*_{\mu\nu}$ the curvature scalar of the
``Einstein metric'' $g^*_{\mu\nu}$ describing the pure spin-2
excitations, and $\varphi$ our long-range scalar field describing
spin-0 excitations. [We use the signature $\scriptstyle -+++$ and
the notation $g_* \equiv -\det g^*_{\mu\nu}$.] The
last term in Eq.~(\ref{eq1.1}) denotes the action of matter,
which is a functional of some matter variables (collectively denoted
by $\psi_m$) and of the ``physical metric'' $\widetilde
g_{\mu\nu}\equiv A^2(\varphi)g^*_{\mu\nu}$. Laboratory clocks and
rods measure the metric $\widetilde g_{\mu\nu}$ which, in the model
considered here, is universally coupled to matter. The reader
will find in Eqs.~(\ref{eq6.1})--(\ref{eq6.7}) below an explicit
example (nonminimally coupled scalar field) of how an action of
the type (\ref{eq1.1}), involving two conformally related metrics
$g^*_{\mu\nu}$ and $\widetilde g_{\mu\nu} = A^2(\varphi)
g^*_{\mu\nu}$, can naturally arise.

The field equations of the theory are most simply formulated in terms
of the pure-spin variables $(g^*_{\mu\nu},\varphi)$. Varying the
action (\ref{eq1.1}) yields
\begin{mathletters}
\label{eq1.2}
\begin{eqnarray}
R^*_{\mu\nu} & = & 2\partial_\mu\varphi \partial_\nu\varphi +
{8\pi G_*\over c^4}\left(T^*_{\mu\nu}-{1\over
2}T^*g^*_{\mu\nu}\right)\ ,
\label{eq1.2a} \\
\Box_{g*}\varphi & = & -{4\pi G_*\over c^4}\alpha(\varphi)T_*\ ,
\label{eq1.2b}\end{eqnarray}
\end{mathletters}
with $T_*^{\mu\nu}\equiv 2 c\, g_*^{-1/2} \delta S_m/\delta
g_{\mu\nu}^*$ denoting the material stress-energy tensor in
``Einstein units'', and
$\alpha(\varphi)$ the logarithmic derivative of $A(\varphi)$~:
\begin{equation}
\alpha(\varphi) \equiv {\partial\ln A(\varphi)\over \partial\varphi}\ .
\label{eq1.3}\end{equation}
[All tensorial operations in Eqs.~(\ref{eq1.2}) are performed by
using the Einstein metric $g_{\mu\nu}^*$, {\it e.g.} $\Box_{g*} \equiv
g_*^{\mu\nu} \nabla_\mu^* \nabla_\nu^*$, $T_*\equiv g^*_{\mu\nu}
T_*^{\mu\nu}$.] As is clear from Eq.~(\ref{eq1.2b}), the quantity
$\alpha(\varphi)$ plays the role of measuring the (field-dependent)
{\it coupling strength\/} between the scalar field and matter. It has
been shown in Refs. \cite{DEF1,DEF2PN} that all {\it weak-field\/}
(``post-Newtonian'') deviations from general relativity (of any
post-Newtonian order) can be expressed in terms of the asymptotic
value of $\alpha(\varphi)$ at spatial infinity and of its successive
scalar-field derivatives. Let $\varphi_0$ denote the asymptotic value
of $\varphi$ at spatial infinity, {\it i.e.}, the ``vacuum
expectation value'' of $\varphi$ far away from the considered
gravitating system. Let us also denote:
$\alpha_0\equiv\alpha(\varphi_0)$,
$\beta_0\equiv\partial\alpha(\varphi_0)/\partial\varphi_0$,
$\beta'_0\equiv\partial\beta(\varphi_0)/\partial\varphi_0$. At the
first post-Newtonian approximation, deviations from general
relativity are proportional to the Eddington parameters
\begin{mathletters}
\label{eq1.4}
\begin{eqnarray}
\overline\gamma & \equiv & \gamma_{\rm Edd}-1 = -2
\alpha_0^2/(1+\alpha_0^2) \ ,
\label{eq1.4a} \\
\overline\beta & \equiv & \beta_{\rm Edd} - 1 = {1\over
2}\beta_0\alpha_0^2/(1+\alpha_0^2)^2 \ ,
\label{eq1.4b}\end{eqnarray}
\end{mathletters}
while at the second post-Newtonian approximation there enters, beyond
$\overline\gamma$ and $\overline\beta$, two new parameters
\cite{DEF1,DEF2PN}
\begin{mathletters}
\label{eq1.5}
\begin{eqnarray}
\varepsilon & = & \beta'_0\alpha_0^3/(1+\alpha_0^2)^3 \ ,
\label{eq1.5a} \\
\zeta & = & \beta_0^2\alpha_0^2/(1+\alpha_0^2)^3 \ .
\label{eq1.5b}\end{eqnarray}
\end{mathletters}
We see explicitly in Eqs.~(\ref{eq1.4}), (\ref{eq1.5}) that all
deviations from general relativity tend to zero with $\alpha_0$ at
least as fast as $\alpha_0^2$. This holds true for {\it weak-field\/}
deviations of arbitrary post-Newtonian order \cite{DEF2PN}. Therefore,
light-deflection or time-delay experiments \cite{gamma} which set
(through Eq.~(\ref{eq1.4a})) the following limit on the coupling
strength of the scalar field,
\begin{equation}
\alpha_0^2 < 10^{-3}\ ,
\label{eq1.6}\end{equation}
tightly constrain the theoretically expectable\footnote{We assume
here the absence of unnaturally large dimensionless numbers appearing
in the successive derivatives of $\alpha(\varphi)$~: $\beta_0$,
$\beta'_0$, \dots} level of deviation from general relativity in all
other experiments probing weak gravitational fields. Note that, in
many physically motivated models, there are much tighter limits on
$\alpha_0^2$ coming from equivalence principle tests (see {\it e.g.}
\cite{DVo} which gets $\alpha_0^2 \lesssim 10^{-7}$ in string-derived
models). These improved limits crucially depend, however, on the
detailed structure and magnitude of equivalence-principle-violating
effects (and disappear in the sub-class of metrically coupled
theories). To stay model-independent, we shall use the
post-Newtonian-derived limit (\ref{eq1.6}) as our standard weak-field
limit. As we shall see later, the importance of the nonperturbative
effects discussed here is not uniformly decreased when $\alpha_0$
takes smaller values, but can level off or even be amplified.

In a previous work \cite{DEF3}, we have shown that experiments
involving the {\it strong gravitational fields\/} of neutron stars
can exhibit a remarkably different behavior from weak-field
solar-system experiments. We proved that when a certain mild
inequality restricting the curvature of the coupling function
$\ln A(\varphi)$ was satisfied, namely
\begin{equation}
\beta_0 \equiv {\partial^2 \ln A(\varphi_0)\over \partial\varphi_0^2}
\lesssim -4\ ,
\label{eq1.7}\end{equation}
nonperturbative strong-gravitational-field effects
developed in neutron stars and induced order-of-unity deviations from
general relativity, even for arbitrary small values of the linear
coupling strength $\alpha_0^2$. The aim of the present paper is to
further study these nonperturbative phenomena and to prepare the
ground for a systematic application to binary-pulsar experiments
\cite{DEFT} by computing the observational effects depending upon the
inertia moments of neutron stars. One of the main results of
the present study will be to show explicitly that binary-pulsar
experiments are, in some regions of theory space, much more
constraining than solar-system experiments. This will be illustrated
in an exclusion plot discussed below.

The organization of this paper is as follows. In Section II,
we show how the non-perturbative scalar-field effects discovered in
\cite{DEF3} can be interpreted as a ``spontaneous scalarization'' of
neutron stars, analogous to the spontaneous magnetization of
ferromagnets. We write in Section III the field equations that must
be numerically integrated to study these non-perturbative effects
in slowly rotating neutron stars. Section IV discusses the
``gravitational form factors'' governing the physics of neutron stars
in tensor--scalar gravity, notably a parameter linked to the
variation of a pulsar's inertia moment caused by the presence of an
orbiting companion. The constraints imposed by three binary-pulsar
experiments on a generic class of tensor--scalar models are then
derived in Section V. Finally, the conclusions of our study are given
in Section VI.

\section{Spontaneous scalarization}
Before tackling the technical problems posed by the computation of
various gravitational ``form factors'' in presence of
strong-scalar-field effects, let us clarify, at the conceptual
level, the physical origin of the nonperturbative effect discovered
in \cite{DEF3}.

Let us consider a very simple coupling function of the form
\begin{equation}
A(\varphi) = A_\beta(\varphi) \equiv
\exp\left({1\over 2}\beta \varphi^2\right)\ ,
\label{eq2.1}\end{equation}
corresponding to a coupling strength $\alpha(\varphi) = \penalty -350
\partial \ln A(\varphi)/\partial\varphi = \beta \varphi$, where
$\beta$ is a given parameter. The model (\ref{eq2.1}), where $\ln
A(\varphi)$ is quadratic in $\varphi$, is second in simplicity to the
Jordan--Fierz--Brans--Dicke model where $\ln A(\varphi) = \alpha_0
\varphi$ is linear in $\varphi$. [We shall sometimes refer to
(\ref{eq2.1}) as ``the quadratic model''.] When $\beta$ satisfies
$\beta\lesssim -4$, we are in a regime where nonperturbative effects
develop for massive enough neutron stars. The results of \cite{DEF3}
raise a paradox in the limit where the asymptotic value of
$\varphi_0$ tends toward zero, {\it i.e.}, $\alpha_0 = \beta
\varphi_0\rightarrow 0$. Indeed, in the case $\alpha = \beta \varphi$
the right-hand side of Eq.~(\ref{eq1.2b}) is proportional to
$\varphi$, and $\varphi(x)\equiv 0$ is an exact solution which
satisfies the homogeneous boundary conditions $\varphi \rightarrow 0$
at spatial infinity. Eq.~(\ref{eq1.2b}) being elliptic in the
stationary case of an isolated star, it would seem that the solution,
with given boundary conditions, must be unique, and therefore that in
the homogeneous case $\varphi_0=0$ the only solution must be the
trivial one $\varphi(x) = 0$. This conclusion is correct in the case
of weakly self-gravitating systems (such as ordinary stars, white
dwarfs or even low-mass neutron stars). Should not then physical
continuity require to take always as ``correct'' solution of
Eq.~(\ref{eq1.2b}) the trivial one, even when considering strongly
self-gravitating systems such as neutron stars~? What can cause a
discontinuity in the configuration of the scalar field (with
homogeneous boundary condition) for massive neutron stars~? In the
simple case of the coupling function (\ref{eq2.1}), we have the
further paradox that the theory is symmetric under the reflection
$\varphi\rightarrow -\varphi$, so that it seems at face value that
the solution of Eqs.~(\ref{eq1.2}) corresponding to the self-symmetric
boundary conditions $\varphi_0 = 0$ must be self-symmetric and
therefore identically zero.

A solution of these paradoxes, and a clearer understanding of the
phenomena studied in \cite{DEF3}, is obtained by making an analogy
with the well-known phenomenon of spontaneous magnetization of
ferromagnets (below the Curie temperature). In the latter case, a
convenient order parameter is the total magnetization ${\bf M}$
(which is thermodynamically conjugate to the external magnetic field
${\bf B}_0$~: ${\bf M}=-\partial E/\partial{\bf B}_0$). In our
``scalarization'' case, we can take as order parameter the total
scalar charge $\omega_A$ developed by the neutron star (labeled $A$)
in presence of an external scalar field $\varphi_0$; it is defined as
the coefficient of $G_*/r$ in the far scalar field around $A$~:
$\varphi(r) = \varphi_0 + G_*\omega_A/r+O(1/r^2)$ as
$r\rightarrow\infty$. As shown in \cite{DEF1}, $\omega_A$ is
energetically conjugate to the external scalar field $\varphi_0$,
\begin{equation}
\omega_A = -\partial m_A/\partial\varphi_0\ ,
\label{eq2.2}\end{equation}
where $m_A$ denotes the total mass-energy of the star (in Einstein
units). It is also the quantity which appears directly in the
Keplerian-order interaction energy between two stars: $V_{\rm int} =
-G_*m_A m_B/r_{AB} - G_*\omega_A\omega_B/r_{AB}$, where the first
term comes from the exchange of a graviton and the second from the
exchange of a scalaron. In the presence of a non-zero external
$\varphi_0$, {\it weakly\/} self-gravitating objects develop a scalar
charge which is proportional to $\varphi_0$ in the limit
$\varphi_0\rightarrow 0$ (``scalar susceptibility''; the analog
to the magnetic susceptibility ${\bf M} = \chi {\bf B}_0$ for
weak external magnetic fields in absence of spontaneous
magnetization).

\begin{figure}
\begin{center}\leavevmode\epsfbox{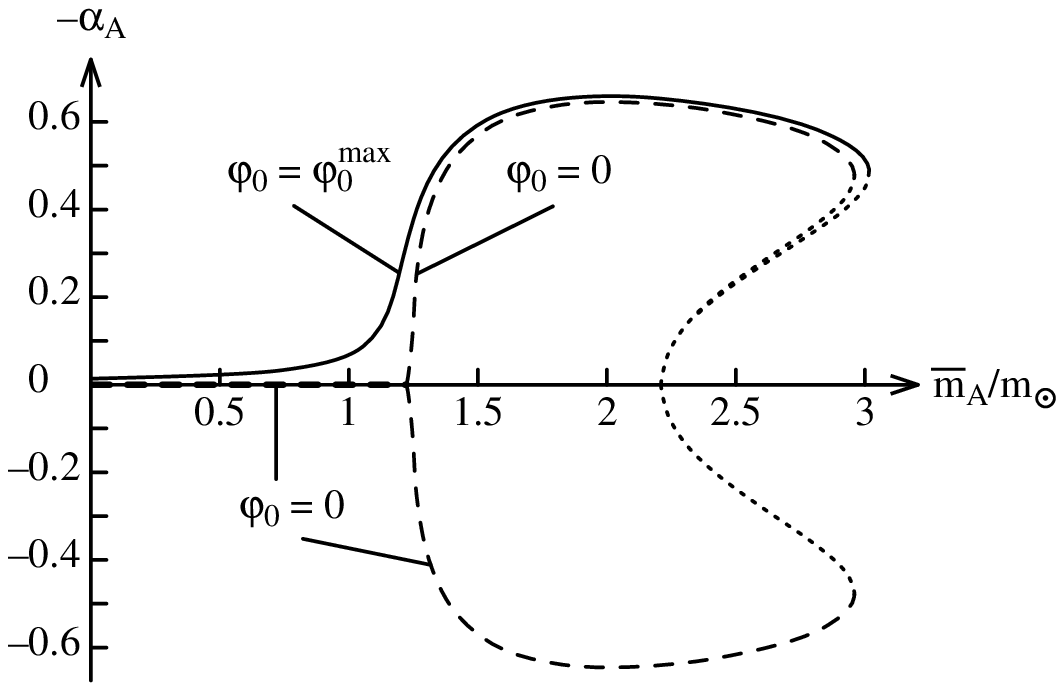}\end{center}\vskip 1pc
\caption{Effective scalar coupling strength $-\alpha_A\equiv
\omega_A/m_A$ versus baryonic mass $\overline m_A$, for the model
$A(\varphi) = \exp(-3 \varphi^2)$. The solid line corresponds to the
maximum value of $\varphi_0$ allowed by solar-system experiments, and
the dashed lines to $\varphi_0 = 0$ (``zero-mode''). The dotted lines
correspond to unstable configurations of the star.}
\label{fig1}
\end{figure}

Following Landau, we can understand what happens for
strongly-self-gravitating objects by writing the total energy to be
minimized as a function of both the external field and the order
parameter, $m_A(\omega_A,\varphi_0) = \mu(\omega_A) -
\omega_A\varphi_0$, and by assuming that the (Legendre transform)
energy function $\mu(\omega_A)$ develops, when some control parameter
varies, a minimum at a non-zero value of $\omega_A$. In our case, if
we fix the shape of the coupling function $A(\varphi)$ (for instance
Eq.~(\ref{eq2.1}) with $\beta$ sufficiently negative), the control
parameter is the total baryon mass $\overline m_A$ of the star. A
simple model exhibiting the appearance of a ``spontaneous
scalarization'' of a star in absence of external field $\varphi_0$ is
simply the usual Landau ansatz near the critical transition point:
$\mu(\omega_A) = {1\over 2} a (\overline m_{\rm cr}-\overline m_A)
\omega_A^2 + {1\over 4} b\, \omega_A^4$. In absence of external field,
$\varphi_0=0$, the energy $m_A$ is minimum at the unique (trivial)
solution $\omega_A=0$ when $\overline m_A < \overline m_{\rm cr}$,
while when $\overline m_A > \overline m_{\rm cr}$, there appear two
energetically favored nontrivial solutions $\omega_A = \pm [b^{-1} a
(\overline m_A - \overline m_{\rm cr})]^{1/2}$. At the critical
transition $\overline m_A = \overline m_{\rm cr}$, the slope
$d\omega_A/d \overline m_A$ is infinite. As in the ferromagnetic case,
the presence of an external field $\varphi_0\neq 0$ {\it smoothes\/}
the transition. For instance, the ``scalar susceptibility'' $\chi_A =
\partial\omega_A/\partial\varphi_0$ which blows up near the critical
point as $|\overline m_A - \overline m_{\rm cr}|^{-1}$ when
$\varphi_0 = 0$ becomes a rapidly varying but smooth function of
$\overline m_A$ when $\varphi_0\neq 0$. The results of \cite{DEF3}
clearly exhibit the sharpening of the transition as
$\varphi_0\rightarrow 0$. This is illustrated in Fig.~\ref{fig1},
which displays two curves corresponding to $\varphi_0 = 2.4\times
10^{-3}$ and $\varphi_0 = 0$ for the same theory ($\beta = -6$ in
Eq.~(\ref{eq2.1})) and the same equation of state (EOS II of Ref.
\cite{DAIC}). Note that, when $\varphi_0\neq 0$, it is the sign of
the external $\varphi_0$ which determines the direction of the
symmetry breaking.

It is convenient, notably for the applications to binary-pulsar
experiments, to replace the quantity $\omega_A$ by the related
quantity
\begin{equation}
\alpha_A \equiv -{\omega_A\over m_A} \equiv {\partial \ln m_A\over
\partial\varphi_0}\ ,
\label{eq2.3}\end{equation}
which measures the effective strength of the coupling between
$\varphi$ and the star. It is the strong-field counterpart of the
weak-field coupling strength $\alpha_0 = \alpha(\varphi_0)$ and
reduces to it in the case of negligible self-gravity. Correlatively,
it is convenient to replace the scalar susceptibility $\chi_A =
\partial\omega_A /
\partial\varphi_0$ by the quantity
\begin{equation}
\beta_A \equiv {\partial\alpha_A\over \partial\varphi_0}\ ,
\label{eq2.4}\end{equation}
which is the strong-field analog of the quantity $\beta_0 =
\partial\alpha(\varphi_0)/\partial\varphi_0$ entering the Eddington
parameter $\beta_{\rm Edd}-1$, Eq.~(\ref{eq1.4b}). The quantity
$\beta_A$ directly enters many observable orbital effects in
binary-pulsar systems \cite{DEF1}.

Summarizing, we conclude that the nonperturbative phenomenon
discussed in \cite{DEF3} can be simply interpreted as a ``spontaneous
scalarization'' phenomenon, {\it i.e.}, a scalar analog of
ferromagnetism. The condition for this phenomenon to occur in actual
neutron stars depends on the equation of state of neutron matter. For
a polytropic model representing a realistic equation of state (with
maximum baryonic mass of $2.23\, m_\odot$ in general relativity), we
found that the critical baryonic mass\footnote{Note that one can
determine the critical baryonic mass as a function of $\beta$, in the
quadratic model (\ref{eq2.1}), by solving a {\it linear\/} problem.
Indeed, the onset of the transition happens when Eq.~(\ref{eq1.2b})
with $\alpha(\varphi)=\beta\varphi$ (and $g_*$ and $T_*$ replaced by a
background general relativistic solution) first admits a
``zero-mode'', {\it i.e.}, a nontrivial homogeneous solution with
vanishing boundary conditions \cite{DEF3}.} for spontaneous
scalarization is smaller than about $1.5\, m_\odot$ (which corresponds
to a general relativistic mass $\approx 1.4\, m_\odot$) when
$\beta_0\equiv \partial^2\ln A(\varphi_0)/\partial\varphi_0^2\leq -5$.
For such values of $\beta_0$, actual neutron stars observed in binary
pulsars would develop strong scalar charges even in absence of
external scalar solicitation ({\it i.e.}, even if $\alpha_0 =
\alpha(\varphi_0) = 0$). For values $-5 \leq \beta_0 \leq -4$, one
can still obtain important deviations from general relativity if the
cosmological value of $\alpha_0$ saturates the present weak-field
limit (\ref{eq1.6}). In all cases, the presence of a non-zero
external $\alpha_0$ smoothes the phase transition and leads to
continuously (but fast) varying values of the effective coupling
parameters $\alpha_A$ and $\beta_A$ as functions of the mass.
Fig.~\ref{fig2} displays the dependence $\overline m_{\rm
cr}(\beta)$ for the quadratic model (\ref{eq2.1}). Some
representative numerical values are quoted in Table~\ref{tab1}.
For $\beta_0$ above some critical value $\beta_{\rm
cr}\approx -4.34$, the maximum mass is reached before the zero-mode
can develop. It is plausible (but difficult to confirm
numerically) that as $\beta \rightarrow \beta_{\rm cr}$,
the critical baryonic mass tends to the general relativistic maximum
baryonic mass ($\approx 2.23\, m_\odot$ in our polytropic model).

The behavior discussed above concerns the scalar models invariant
under the reflection symmetry $\varphi \rightarrow - \varphi$, such as
$A(\varphi) = \exp({1\over 2}\beta \varphi^2)$ or $A(\varphi) =
\cos(\sqrt{-\beta}\varphi$). A dissymmetric coupling function, such as
$A(\varphi) = \exp({1\over 2}\beta\varphi^2 + {1\over
6}\beta'\varphi^3)$ would lead to hysteresis phenomena (first-order
rather than second-order phase transition): for some values of the
control parameter $\overline m_A$, there will be two locally stable
energy minima available. The scalar configuration chosen by the star
would depend on the route taken to evolve into its present mass
state. Let us also mention that we would get an even richer
(Goldstone-like) phenomenology if we were to consider models involving
several scalar fields, with {\it e.g.} spontaneous breaking of a
continuous symmetry in the scalar-field space. Finally, let us make
it clear that a negative value for $\beta_0 \equiv \partial^2 \ln
A/\partial\varphi_0^2$ does not mean at all that we are introducing
some pathology in our scalar-field model. The theories we consider
are well-behaved field models having only positive-energy
excitations. A negative value of $\beta_0$ means only that scalar
field nonlinearities can reinforce the naturally attractive
character of scalar interactions, so that it becomes energetically
favorable to generate more scalar-field energy\footnote{The
appearance of a negative critical value of $\beta_0$ can be easily
understood in the lowest approximation, where the scalar energy
functional to be minimized reads (when setting $G=c=1$):
$E[\varphi] = \int d^3x\left[{1\over 8\pi} (\partial_i \varphi)^2 +
\rho(1+{1\over 2} \beta\varphi^2)\right]$. Indeed, let us consider
for instance the simplest trial continuous field configurations,
$\varphi(r) = {\rm const.} = \omega_A/R$ inside a star of mass $m =
\int d^3x\,\rho$, and $\varphi(r)=\omega_A/r$ outside the star
($r>R$). This yields $E[\omega_A] = m+{1\over 2} C \omega_A^2$, where
$C = R^{-1}(1+\beta m/R)$ becomes negative for a sufficiently
negative $\beta$. The missing stabilizing contribution $+{1\over 4}
b \omega_A^4$ would come from taking into account higher-order
nonlinearities.}.

\section{Slowly rotating neutron stars in tensor--scalar gravity}
One of the main objects of the present paper is to show how to
compute the moments of inertia of slowly rotating neutron stars in
tensor--scalar gravity, especially in presence of the nonperturbative
strong-scalar-field effects recalled above. We shall work in the
Einstein conformal frame, within which the basic global mechanical
quantities, such as total mass and total angular momentum, are
conserved (in absence of radiation or particle exchange) and can, as
usual, be read off the asymptotic expansion of the metric. The total
mass $m_A$ (in Einstein units) can be read off the $1/r$ behavior of
$g_{00}^*$ or $g_{ij}^*$, while the $z$-component of the total angular
momentum $J_A$ (in Einstein units) can be read off the $1/r^2$
behavior of the mixed component $g_{0i}^*$. We consider only
stationary axisymmetric field configurations. It has been shown by
Hartle \cite{H67} (see also \cite{C87}) that the metric corresponding
to a slowly rotating star could be written, when keeping only first
order terms in the angular velocity $\Omega = u^\phi/u^t$, as
\begin{eqnarray}
&& ds_*^2 = g_{\mu\nu}^* dx^\mu dx^\nu = - e^{\nu(\rho)} c^2 dt^2 +
e^{\mu(\rho)} d\rho^2
\nonumber \\
&& + \rho^2 d\theta^2 + \rho^2 \sin^2\theta \Bigl(
d\phi+\bigl[\omega(\rho,\theta)-\Omega\bigr] dt \Bigr)^2\ .
\label{eq3.1}\end{eqnarray}
Thanks to the neglect of fractional corrections of order $\Omega^2$,
the diagonal metric coefficients $\nu(\rho)$ and $\mu(\rho)$ can be
taken to be the solutions corresponding to a spherically symmetric
non-rotating star. The only new field variable which appears in the
slowly rotating case is the function $\omega(\rho,\theta)$ entering
the mixed component $g^*_{t\phi}=\rho^2\sin^2\theta
[\omega(\rho,\theta)-\Omega]$. The subtraction of the star's angular
velocity $\Omega$ is chosen for later convenience\footnote{With this
definition of variables, the stress-energy tensor of the fluid gives
simply $T_{*\phi}^t = (\epsilon_*+p_*) e^{-\nu} \rho^2 \omega
\sin^2\theta$ thanks to a combination between $g^*_{t\phi}$ and
$\Omega = u^\phi/u^t$.}. The total angular momentum $J_A$ is read
off the $1/\rho^3$ behavior of $\omega$~:
\begin{equation}
\omega = \Omega - {G_*\over c^2}\, {2J_A\over \rho^3} +O\left({1\over
\rho^4} \right)\ .
\label{eq3.2}\end{equation}
Then the inertia moment (in Einstein units) is defined, in the slow
rotation limit, as the ratio
\begin{equation}
I_A = {J_A\over \Omega} +O(\Omega^2)\ .
\label{eq3.3}\end{equation}
We need now to write down explicitly the field equations (\ref{eq1.2}).
As the scalar field $\varphi$ does not couple linearly to the rotation,
the field equation for $\varphi$ is, modulo terms of order $\Omega^2$,
the same as for a spherically symmetric, non-rotating star (therefore
$\varphi$ will, modulo $O(\Omega^2)$, be spherically symmetric). The
field equation for the new variable $\omega$ comes from
\begin{equation}
R_{*\phi}^t = 2 \partial_\phi\varphi g_*^{t\alpha}
\partial_\alpha\varphi
+{8\pi G_*\over c^4}T_{*\phi}^t\ .
\label{eq3.4}\end{equation}
Simply from axisymmetry ($\partial_\phi = 0$) we see that the scalar
contribution to the right-hand side of (\ref{eq3.4}) vanishes exactly.
We are then left with the usual Einstein field equations with a
localized material source. Taking as usual a perfect fluid description
of nuclear matter (with energy density $\epsilon_*$ and pressure $p_*$
in Einstein units) we can directly use the results of Refs.
\cite{H67,C87}. [One must, however, be careful not to use equations
where the ``diagonal'' Einstein field equations have been replaced.]
We find the following homogeneous equation for $\omega$~:
\begin{eqnarray}
&& {1\over \rho^4}\partial_\rho\left[
\rho^4 e^{-(\nu+\mu)/2} \partial_\rho\omega
\right]
+{e^{(\mu-\nu)/2}\over \rho^2\sin^3\theta}\partial_\theta\left(
\sin^3\theta \partial_\theta\omega
\right)
\nonumber\\
&& = {16\pi G_*\over c^4}(\epsilon_*+p_*)e^{(\mu-\nu)/2}\omega\ .
\label{eq3.5}
\end{eqnarray}
As in Refs. \cite{H67,C87}, a decomposition of $\omega(\rho,\theta)$
in associated Legendre polynomials $d P_\ell(\cos\theta)/d\cos\theta$
shows that there is only a $P$ contribution ($\ell=1$), so that, in
fact, $\omega$ depends only on $\rho$ and not on $\theta$. Adding the
scalar-modified diagonal Einstein equations (written in
\cite{DEF3}), we finally get the following complete set of radial
equations for our field variables (a prime denoting $d/d\rho$):
\widetext
\begin{mathletters}
\label{eq3.6}
\begin{eqnarray}
M' & = & {4\pi G_*\over c^4} \rho^2 A^4(\varphi) \widetilde\epsilon
+{1\over 2}\rho(\rho-2M)\psi^2\ ,
\label{eq3.6a} \\
\nu' & = & {8\pi G_*\over c^4}\, {\rho^2 A^4(\varphi)\widetilde p\over
\rho-2M} + \rho\psi^2 +{2M\over \rho(\rho-2M)}\ ,
\label{eq3.6b} \\
\varphi' & = & \psi\ ,
\label{eq3.6c} \\
\psi' & = & {4\pi G_*\over c^4}\, {\rho A^4(\varphi)\over \rho-2M}
\left[\alpha(\varphi)(\widetilde\epsilon-3\widetilde p) + \rho\psi
(\widetilde\epsilon-\widetilde p)
\right]
-{2(\rho-M)\over \rho(\rho-2M)}\psi\ ,
\label{eq3.6d} \\
\widetilde p' & = & -(\widetilde\epsilon+\widetilde p)\left[
{4\pi G_*\over c^4}\, {\rho^2 A^4(\varphi)\widetilde p\over \rho- 2M}
+{1\over 2}\rho\psi^2 + {M\over \rho(\rho-2M)}+\alpha(\varphi)\psi
\right]\ ,
\label{eq3.6e} \\
\overline M' & = & 4\pi\widetilde m_b \widetilde n A^3(\varphi)
{\rho^2\over\sqrt{1-2M/\rho}}\ ,
\label{eq3.6f} \\
\omega' & = & \varpi\ ,
\label{eq3.6g} \\
\varpi' & = & {4\pi G_*\over c^4}\, {\rho^2\over \rho-2M}A^4(\varphi)
\left(\widetilde\epsilon+\widetilde p\right)\left(
\varpi + {4\omega\over \rho}\right)
+\left(\psi^2\rho - {4\over \rho}\right)\varpi\ .
\label{eq3.6h}\end{eqnarray}
\end{mathletters}
\narrowtext
The notation used in Eqs.~(\ref{eq3.6}) is the following: $M(\rho)$ is
defined by writing the radial metric coefficient $g_{\rho\rho}$ as
$e^{\mu(\rho)}\equiv (1-2M(\rho)/\rho)^{-1}$. As usual the value of
$M(\rho)$ at infinity is the total (ADM) mass. The fluid variables have
been expressed in physical units using $T_{*\mu}^\nu =
A^4(\varphi)\widetilde T_\mu^\nu$. [It is in these units that one can
write a usual equation of state $\widetilde\epsilon =
\widetilde\epsilon(\widetilde n)$, $\widetilde p = \widetilde
p(\widetilde n)$, where $\widetilde n$ denotes the physical number
density of baryons.] $\psi$ and $\varpi$ are just intermediate
notations for the radial derivatives of $\varphi$ and $\omega$,
respectively. Finally, we have added an equation for the radial
distribution of the baryonic mass $\overline m_A = \overline M(R) =
\widetilde m_b\int_A \widetilde n \sqrt{\widetilde g}\,
\widetilde u^0 d^3 x = \widetilde m_b \int_0^R 4\pi\widetilde n
A^3(\varphi) \rho^2 (1-2M/\rho)^{-1/2} d\rho$, where $R$ denotes
the (Schwarzschild-coordinates) radius of the star ({\it i.e.},
the value of $\rho$ where $\widetilde p$ and $\widetilde\epsilon$
vanish).

Note that several of the right-hand sides of Eqs.~(\ref{eq3.6})
contain terms proportional to $\psi^2 = (\varphi')^2$ ({\it i.e.},
proportional to the scalar-field energy density). These terms do not
vanish outside the star. However, one can avoid numerically
integrating Eqs.~(\ref{eq3.6}) up to $\rho=\infty$ by matching the
result of integrating (\ref{eq3.6}) up to the radius $R$ of the star
to the known general form of the exact static, spherically symmetric
exterior solution. This is, however, a bit subtle because the general
exterior solution can only be written in closed form in some
special coordinates introduced by Just \cite{J59,CEF90,DEF1}, or
(through a simple transformation) in isotropic coordinates, but not
in the Schwarzschild coordinates we are using. Still, it was shown in
\cite{DEF3} how to extract, via a matching across the star's surface,
the global quantities $m_A$ and $\alpha_A$ from the results of
integrating Eqs.~(\ref{eq3.6}) up to $\rho = R$. We need to do here
more work to extract $J_A$ (and $I_A$) from the results for the
variables $\omega$ and $\varpi \equiv d\omega/ d\rho$.

Outside the star, Eq.~(\ref{eq3.5}) (with $\partial_\theta=0$) shows
directly that $\rho^4 e^{-(\nu+\mu)/2}\partial_\rho\omega$ is a
constant. From Eq.~(\ref{eq3.2}), this constant is simply related to
the total angular momentum, so that
\begin{equation}
{d\omega\over d \rho} = 6{G_*\over c^2} J_A {e^{(\nu+\mu)/2}\over
\rho^4} \quad\mbox{(outside the star)}.
\label{eq3.7}\end{equation}
Eq.~(\ref{eq3.7}) gives one equation to determine $J_A$. We need
another equation to determine $\Omega$ and then $I_A \equiv
J_A/\Omega$. Note that the equation for $\omega$ ({\it e.g.}
Eq.~(\ref{eq3.5})) is homogeneous in $\omega$. Therefore, we can
start the radial integration with an arbitrary (non zero) value of
$\omega(\rho)$ at $\rho=0$, but we need to extract from
$\omega(\rho)$ the value of the fluid angular velocity $\Omega$
implied by this arbitrary choice. To achieve this, it suffices to
integrate explicitly Eq.~(\ref{eq3.7}) with the boundary condition
$\omega(\rho) \rightarrow \Omega$ when $\rho\rightarrow\infty$ (as is
clear from Eqs.~(\ref{eq3.1}) or (\ref{eq3.2})). This integration can
be done by rewriting Eq.~(\ref{eq3.7}) in Just radial coordinate $r$.
Indeed, the general exterior static, spherically symmetric solution
\cite{J59,CEF90,DEF1} reads
\begin{mathletters}
\label{eq3.8}
\begin{eqnarray}
ds_*^2 & = & -e^\nu c^2 dt^2
\nonumber\\
&& + e^{-\nu}\left[
dr^2 + (r^2-ar)(d\theta^2+\sin^2 \theta d\phi^2)
\right]\, ,
\label{eq3.8a} \\
e^{\nu(r)} & = & \left(
1-{a\over r}
\right)^{b/a}\ ,
\label{eq3.8b} \\
\varphi(r) & = & \varphi_0 + {d\over a} \ln\left(
1-{a\over r}\right)\ ,
\label{eq3.8c}\end{eqnarray}
\end{mathletters}
where the integration constants $a,b,d$ are constrained by $a^2-b^2=4
d^2$, and are expressible in terms of the total Einstein mass $m_A$ and
the effective coupling constant $\alpha_A$, Eq.~(\ref{eq2.3}), via
\begin{mathletters}
\label{eq3.9}
\begin{eqnarray}
b & = & 2{G_*\over c^2}m_A\ ,
\label{eq3.9a} \\
{a\over b} & = & \sqrt{1+\alpha_A^2}\ ,
\label{eq3.9b} \\
{d\over b} & = & {1\over 2}\alpha_A\ .
\label{eq3.9c}\end{eqnarray}
\end{mathletters}
Comparing Eq.~(\ref{eq3.8a}) with the Schwarzschild form (\ref{eq3.1})
yields
\begin{mathletters}
\label{eq3.10}
\begin{eqnarray}
\rho & = & r\left(
1-{a\over r}
\right)^{(a-b)/2a}\ ,
\label{eq3.10a} \\
e^\mu & = & \left(
1-{a\over r}
\right)\left(
1-{a+b\over 2r}
\right)^{-2}\ .
\label{eq3.10b}\end{eqnarray}
\end{mathletters}
Inserting these results into Eq.~(\ref{eq3.7}) leads to an elementary
integral for $\omega(r)$. To write explicitly the answer it is
convenient to introduce the parameter
\begin{equation}
p\equiv {1\over a}\ln \left(
1-{a\over r}
\right)\ .
\label{eq3.11}\end{equation}
In terms of $p$, the exact exterior solution for $\omega$ reads
\begin{eqnarray}
\omega &=& \Omega + {6 G_* J_A\over c^2 b (4b^2-a^2)}\Biggl\{
e^{2bp} - 1 + e^{2bp}\times
\nonumber\\
&& \times\left[
\left({2b\over a}\right)^2 \left(\cosh(ap)-1\right)
-{2b\over a} \sinh(ap)
\right]
\Biggr\}\ .
\label{eq3.12}\end{eqnarray}
Combining the results just derived on the radial dependence of $\omega$
with the results of \cite{DEF3} for the matching of the other field
variables, we can finally write a set of equations allowing one to
extract all the needed physical quantities from the surface values
obtained from integrating Eqs.~(\ref{eq3.6}) from the center $\rho =
0$~:
\begin{mathletters}
\label{eq3.13}
\begin{eqnarray}
R & \equiv & \rho_s\ ,
\label{eq3.13a} \\
\nu'_s & \equiv & R\psi_s^2 +{2M_s\over R(R-2M_s)}\ ,
\label{eq3.13b} \\
\alpha_A & \equiv & 2\psi_s\over \nu'_s\ ,
\label{eq3.13c} \\
Q_1 & \equiv & \left(1+\alpha_A^2\right)^{1/2}\ ,
\label{eq3.13d} \\
Q_2 & \equiv & \left(1-2M_s/R\right)^{1/2}\ ,
\label{eq3.13e} \\
\widehat\nu_s & \equiv & -{2\over Q_1}\,
{\rm arctanh}\left({Q_1\over 1+2
(R\nu'_s)^{-1}}\right)\ ,
\label{eq3.13f} \\
\varphi_0 & \equiv & \varphi_s - {1\over 2}\alpha_A \widehat\nu_s\ ,
\label{eq3.13g} \\
{G_*\over c^2}m_A & \equiv & {1\over 2}\nu'_s R^2 Q_2
\exp\left({1\over 2}\widehat\nu_s\right)\ ,
\label{eq3.13h} \\
\overline m_A & \equiv & \overline M_s\ ,
\label{eq3.13i} \\
{G_*\over c^2} J_A & \equiv & {1\over 6}\varpi_s R^4 Q_2
\exp\left(-{1\over 2}\widehat\nu_s\right)\ ,
\label{eq3.13j} \\
\Omega & \equiv & \omega_s - {c^4\over G_*^2}\,
{3 J_A\over 4 m_A^3(3-\alpha_A^2)}\,
\biggl\{
e^{2\hat\nu_s} - 1 + {4 G_*m_A\over
Rc^2}\times
\nonumber\\
&& \times e^{\hat\nu_s}
\left[
{2 G_*m_A\over Rc^2} + e^{\hat\nu_s/2}
\cosh\left({1\over 2}Q_1 \widehat \nu_s\right)
\right]
\biggr\}\ ,
\nonumber\\
&& \label{eq3.13k} \\
I_A & \equiv & {J_A\over \Omega}\ .
\label{eq3.13l}\end{eqnarray}
\end{mathletters}
The notation used in Eqs.~(\ref{eq3.13}) is that a suffix $s$ denotes
the surface value of any of the variables entering the first-order
system (\ref{eq3.6}). The only exception (apart from $\nu'_s$ that we
redefine explicitly as the surface value of the right-hand side of
Eq.~(\ref{eq3.6b})) is $\widehat\nu_s$, which is the ``correct'' value
of $\nu$ at the surface when $\nu$ is normalized as being zero at
infinity. Indeed, as the system (\ref{eq3.6}) is integrated from the
center (starting with an arbitrary value of $\nu(0)$) up to the
surface, the surface value of $\nu(\rho)$ naively obtained from
integrating (\ref{eq3.6}) is not the one to be used in any of the
physically normalized results.

Let us finally mention the set of initial conditions, at the center,
used for integrating Eqs.~(\ref{eq3.6}). Actually, because of the
singular nature of the point $\rho=0$, one numerically imposes initial
conditions at a small but nonzero radius $\rho_{\rm min}$. The values
of some of the radial derivatives ($\varphi'\equiv\psi$ and
$\omega'\equiv\varpi$) are determined so as to be consistent with
regular Taylor expansions at the origin (for instance, writing
$\varphi(\rho) = \varphi({\bf x}) = \varphi({\bf 0}) + {1\over 6}
{\bf x}^2 \Delta\varphi({\bf 0}) + O({\bf x}^4)$ determines
$\varphi'(\rho)\sim {1\over 3} \rho \Delta\varphi(0)$ as
$\rho\rightarrow 0$). The complete set of initial conditions reads:
\begin{mathletters}
\label{eq3.14}
\begin{eqnarray}
M(\rho_{\rm min}) & = & 0\ ,
\label{eq3.14a} \\
\nu(\rho_{\rm min}) & = & 0\ ,
\label{eq3.14b} \\
\varphi(\rho_{\rm min}) & = & \varphi_c\ ,
\label{eq3.14c} \\
\psi(\rho_{\rm min}) & = & \left(
{1\over 3} \rho_{\rm min}\right)
\times {4\pi G_*\over c^4}A^4(\varphi_c)
\nonumber\\
&& \times \alpha(\varphi_c)
\left[
\widetilde\epsilon(\widetilde n_c)
-3 \widetilde p(\widetilde n_c)\right]\ ,
\label{eq3.14d} \\
\widetilde n(\rho_{\rm min}) & = & \widetilde n_c\ ,
\label{eq3.14e} \\
\overline M(\rho_{\rm min}) & = & 0\ ,
\label{eq3.14f} \\
\omega(\rho_{\rm min}) & = & 1\ ,
\label{eq3.14g} \\
\varpi(\rho_{\rm min}) & = & \left(
{1\over 5} \rho_{\rm min}\right)
\times {16\pi G_*\over c^4}A^4(\varphi_c)
\nonumber\\
&& \times \left[
\widetilde\epsilon(\widetilde n_c)
+ \widetilde p(\widetilde n_c)\right] \omega(\rho_{\rm min})\ .
\label{eq3.14h}\end{eqnarray}
\end{mathletters}
Note that (as discussed above) the initial conditions (\ref{eq3.14b})
and (\ref{eq3.14g}) are arbitrary, and that we transform
Eq.~(\ref{eq3.6e}) in an evolution equation for the physical number
density $\widetilde n$ using the equation of state, {\it i.e.},
$\widetilde p^{\,\prime} = (d\widetilde p(\widetilde n)/d \widetilde
n)\times \widetilde n'$. The choice of $\varphi_c$ and $\widetilde
n_c$ is discussed below.

\section{The gravitational form factors of rotating neutron stars}
\subsection{Scalar-field dependence of the inertia moment}
Extending the analysis of \cite{DEF3}, we have studied the impact of
scalar-induced strong-field effects on the gravitational form factors
of neutron stars. By ``gravitational form factor'' we mean the set of
coupling constants that appear, within tensor--scalar theories, in the
description of the relativistic motion and timing of binary (and
isolated) pulsars. As discussed in detail in \cite{DEF1}\footnote{We
restrict here the more general results of \cite{DEF1} to the simple
case where there is only one scalar field.}, the $(v/c)^2$-accurate
orbital dynamics of binary systems depends, besides the Einstein
masses of the two objects $m_A$ and $m_B$, on the effective scalar
coupling constants $\alpha_A$, $\alpha_B$, defined in
Eq.~(\ref{eq2.3}), as well as on their scalar-field derivatives
$\beta_A$, $\beta_B$, Eq.~(\ref{eq2.4}). It was also shown in
\cite{DEF1} that the same parameters $\alpha_A, \alpha_B, \beta_A,
\beta_B$, suffice to express all radiation reaction effects (up to
$O(v^7/c^7)$) in a tensor--scalar description of compact binary
systems. On the other hand, the relativistic timing of binary-pulsar
systems involves, besides the above $\alpha$'s and $\beta$'s, a new
parameter describing the possible field dependence of the inertia
moment $I_A$ of the pulsar. [In the following, we use the label $A$
to indicate the timed pulsar, by opposition to the companion labeled
$B$.] Indeed, as pointed out by Eardley \cite{E75} (see also
\cite{W93}), the adiabatic invariance (under the slow variation of
the local scalar-field environment caused by the motion of the
companion) of the total angular momentum of the pulsar $J_A =
I_A(\varphi_{0A}^{\rm loc})\Omega_A$ implies that the angular
velocity of the pulsar $\Omega_A$ will fluctuate in response to the
orbital-induced variations of the external scalar field
$\varphi_{0A}^{\rm loc}$ locally felt by the pulsar. As discussed in
more detail below, the observable deviations from general relativity
implied by this effect are given by the parameter $K_A^B \equiv
-\alpha_B \partial\ln I_A/\partial\varphi_0$, in which $I_A$ denotes,
as above, the inertia moment of the pulsar in (local) Einstein units.

To compute $\partial \ln I_A/\partial\varphi_0$, we have numerically
integrated equations (\ref{eq3.6}) with a suitable ``shooting''
strategy for the choice of initial conditions. Indeed, the quantities
that are physically fixed are $\varphi_0$ (the value of
$\varphi$ far from the star) and $\overline m_A$ (the baryonic mass
of the neutron star). [Note that when a derivative with respect to
$\varphi_0$ is taken, as in the definitions of $\beta_A$,
Eq.~(\ref{eq2.4}), or of $K_A^B$, it must be performed for a
fixed value of $\overline m_A$.] Therefore, by trial and error, one
must vary the initial conditions $\varphi_c$ and
$\widetilde n_c$ in Eqs.~(\ref{eq3.14}) until they lead to the desired
values of $\varphi_0$ and $\overline m_A$. In the end, one wants to
explore the way the observables $m_A, \alpha_A, \beta_A, I_A,
\partial\ln I_A/ \partial\varphi_0, \ldots$ depend upon $\varphi_0$
and $\overline m_A$.

The values of $m_A, \alpha_A, \ldots$ as functions of $\varphi_0$ and
$\overline m_A$ depend upon the equation of state used to describe
the nuclear matter in the neutron star. We shall discuss in a later
publication the dependence of our results on the choice of the
equation of state. In the present work, we shall consider, for
simplicity, only a fixed polytropic equation of state:
\begin{mathletters}
\label{eq4.1}
\begin{eqnarray}
\widetilde \epsilon & = & \widetilde n \widetilde m_b +
{K \widetilde n_0 \widetilde m_b\over \Gamma - 1}
\left({\widetilde n\over \widetilde n_0}\right)^\Gamma \ ,
\label{eq4.1a} \\
\widetilde p & = & K \widetilde n_0 \widetilde m_b \left({\widetilde
n\over \widetilde n_0}\right)^\Gamma\ .
\label{eq4.1b}\end{eqnarray}
\end{mathletters}
All quantities in Eqs.~(\ref{eq4.1}) are in local physical units;
$\widetilde m_b \equiv 1.66\times 10^{-24}\ {\rm g}$ is a fiducial
baryon mass and $\widetilde n_0\equiv 0.1\ {\rm fm}^{-3}$ a typical
nuclear number density. We shall use the following specific values of
the polytropic parameters $\Gamma$ and $K$,
\begin{equation}
\Gamma = 2.34 \quad,\quad K=0.0195\ ,
\label{eq4.2}\end{equation}
which have been chosen to fit a realistic equation of state which is
neither too hard nor too soft: the equation of state II of Ref.
\cite{DAIC}. [The polytropic constant $K$ should not be
confused with the parameter $K_A^B$ linked to the
scalar-field-induced variation of the inertia moment.] The precise
values (\ref{eq4.2}) were adjusted to fit the curve giving, in
general relativity, the fractional binding energy $f\equiv (\overline
m - m)/m$ as a function of the baryonic mass. In particular they lead
to the same maximum baryonic mass, $\overline m_{\rm max} = 2.23\,
m_\odot$, in general relativity. Let us note in passing that to
convert from the nuclear fiducial quantities to more adequate
astrophysical units ($m_\odot$ for masses, $G_*m_\odot/c^2$ for
distances), it is convenient to use the numerical value
\begin{equation}
{4\pi G_* \widetilde n_0 \widetilde m_b\over c^2}\left(
{G_* m_\odot\over c^2}\right)^2
= {1\over 296.135}\ .
\label{eq4.3}\end{equation}
For technical convenience, when comparing different theories we keep
fixed $G_* = 6.67\times 10^{-8} {\rm cm}^3 {\rm g}^{-1} {\rm s}^{-2}$
(and $m_\odot = 1.99 \times 10^{33} {\rm g}$, measured in $g^*$
units). See Ref.~\cite{DEF1} for the factors (differing from unity by
$\lesssim 10^{-3}$) relating $g^*$-frame quantities to directly
observable ones.

We present in Figure \ref{fig3} some of our numerical results for the
dependence upon the baryonic mass of $\alpha_A$,
$\beta_A$, $I_A$ [in units of $m_\odot(G_* m_\odot/c^2)^2$] and
$\partial\ln I_A/\partial\varphi_0$. All the results of these Figures
have been computed within the tensor--scalar theory defined by the
particular coupling function
\begin{equation}
A_{\overline 6}(\varphi) \equiv \exp(-3\varphi^2)\ .
\label{eq4.4}\end{equation}
This model belongs to the class of quadratic models (\ref{eq2.1}),
and possesses a curvature parameter for the logarithm of the coupling
function, $\beta = \beta_0 = \partial^2\ln A/\partial\varphi_0^2 =
-6$. In the limit where $\varphi_0\rightarrow 0$, this model exhibits
a spontaneous scalarization above a critical baryonic mass
$\overline m_{\rm cr} = 1.24\, m_\odot$. As explained in Section II,
the presence of a nonzero external scalar background
$\varphi_0\neq 0$ smoothes the scalarization and leads to continuous
variations of $\alpha_A, \beta_A, \ldots$ in function of $\overline
m_A$. For instance, instead of having a Curie-type blow-up $\propto
|\overline m_A - \overline m_{\rm cr}|^{-1}$ for the
zero-external-field ``susceptibility'' $\beta_A = \partial
\alpha_A/\partial\varphi_0$, we get a ``resonance'' bump in
$\beta_A$ when $\overline m_A \approx \overline m_{\rm cr}$. There
remains however an infinite blow-up in $\beta_A$ when $\overline m_A$
reaches the maximum baryonic mass. It is easy to see analytically
that this blow-up must be there. (The same remark applies to
$\partial\ln I_A/\partial\varphi_0$.) For definiteness, we have drawn
Fig. \ref{fig3} for the value
\begin{equation}
\varphi_0 = \varphi_0^{\rm max} \equiv 2.4 \times 10^{-3}\ ,
\label{eq4.5}\end{equation}
which is the maximum value of $\varphi_0$ allowed by present
weak-field tests within the model (\ref{eq4.4}). This maximum value
is obtained from considering not only the limit $\alpha_0^2 <
10^{-3}$, Eq.~(\ref{eq1.6}), coming from time-delay and
light-deflection experiments
\cite{gamma}, but also the limit
\begin{equation}
|\beta_0| \alpha_0^2 < 1.2\times 10^{-3}
\label{eq4.6}\end{equation}
coming from the lunar-laser-ranging constraint $|\overline\beta| <
6\times 10^{-4}$ \cite{LLR} on the Eddington parameter $\overline
\beta \equiv \beta_{\rm Edd} -1\approx {1\over 2} \beta_0 \alpha_0^2$
(see Eq. (\ref{eq1.4b})). When $|\beta_0|>1.2$, the limit
(\ref{eq4.6}) is more stringent than (\ref{eq1.6}) and defines the
maximal allowed value for $|\alpha_0|$ and thereby for $|\varphi_0|
\approx |\alpha_0/\beta_0|$ (see the exclusion plot in section
V.D. below).

Besides the variation of the shapes of the curves in Fig. \ref{fig3}
when $\varphi_0$ is allowed to vary (which is always a sharpening of
the bumps and a stabilization of the other features\footnote{See, for
instance, Fig. \ref{fig1} above which shows that the wide plateau in
$\alpha_A$, beyond $\overline m_{\rm cr}$, varies very little when
$\varphi_0$ tends to zero.}), we have also numerically explored the
effect of varying the curvature parameter $\beta$ in Eq.~(\ref{eq2.1}).
The two main effects of varying $\beta$ are (i)~to enlarge the
values of the form factors $|\alpha_A|$, $|\beta_A|$, $|\partial\ln
I_A/\partial\varphi_0|$ as $-\beta$ increases, and (ii)~to displace
the location of the critical point $\overline m_{\rm cr}$. For
instance, we find (within the models (\ref{eq2.1})) $\overline
m_{\rm cr}(\beta= -5) = 1.56\, m_\odot$, $\overline m_{\rm cr}(\beta=
-4.5) = 1.84\, m_\odot$. These values are below the (expected) maximum
mass of a neutron star. However, observed neutron stars have baryonic
masses around $1.5\, m_\odot$ (corresponding to general relativistic
Einstein masses around $1.4\, m_\odot$), therefore we expect that
strong-scalar-field effects can have significant observational
consequences only when $\beta\leq -5$.

\subsection{Scalar-field effects in the timing parameter $\gamma$}
Up to now, the non-Einsteinian effects linked to the field
dependence of the inertia moment have been treated by an
approximation \cite{E75,W93,DEF1} which is insufficient for tackling
the nonperturbative phenomena discussed here. One of the main aims
of the present paper is to remedy this situation. Let us first
clarify the observable effect of the variation of the pulsar
inertia moment with the local scalar background\footnote{This denotes
the nearly uniform value of $\varphi$ on a sphere centered on $A$
having a radius much larger than the radius of the neutron star $A$
but much smaller than the distance to the companion.}
$\varphi_A\equiv \varphi_{0A}^{\rm loc}$ \cite{E75,W93}.

The central tool of binary-pulsar experiments is the ``timing
formula'' (see {\it e.g.} \cite{DD86,DT92}), {\it i.e.}, the
mathematical function relating the ``intrinsic time'' of the pulsar
clock $T$ to the arrival time on Earth of radio pulses. The
successive ticks of the pulsar time $T$ are defined to correspond to
successive $2\pi$ rotations of the pulsar around itself: $\phi^{\rm
PSR} = 2\pi T/P_p$, where $P_p$ is the intrinsic period of the pulsar
(for simplicity we neglect here the slow-down of the rotation of the
pulsar as well as aberration effects). In other words, adding the
label $A$ and passing to a differential formulation, $d T_A = C
d\phi_A$ for a certain constant $C$. In (local) Einstein units, the
pulsar angular momentum reads $J_A = I_A \Omega_A = I_A
d\phi_A/d\tau^*_A$, where $d\tau^*_A = |ds^*_A|/c =
(-g_{\mu\nu}^{*A}dz_A^\mu dz_A^\nu)^{1/2}/c$ is the Einstein proper
time in a local inertial frame around $A$. The angular momentum $J_A$
is an action variable ($J = p_\phi = {1\over 2\pi} \oint p_i dq^i$)
and therefore an adiabatic invariant under slow changes of
parameters. It remains therefore constant as the pulsar moves on its
orbit and feels a slowly changing $\varphi_A$ from its companion.
This yields $dT_A = C' d\tau_A^*/I_A$ for some new constant $C'$. The
latter equation can be approximately rewritten in terms of some
coordinate time $t$ used to describe the binary motion:
\begin{equation}
dT_A \approx C' \sqrt{-g^{*A}_{00}}\sqrt{1-{\bf v}_A^2/c^2}\,
dt/I_A\left(\varphi_A(t)\right)\ ,
\label{eq4.7}\end{equation}
where (to sufficient accuracy) ${\bf v}_A^2$ is the Euclidean square of
the coordinate velocity of the pulsar ${\bf v}_A = d {\bf z}_A/dt$.
Using (see \cite{DEF1})
\begin{mathletters}
\label{eq4.8}
\begin{eqnarray}
\sqrt{-g^{*A}_{00}} & = & 1 - {G_* m_B\over r_{AB}c^2} +O\left({1\over
c^4}\right)\ ,
\label{eq4.8a} \\
\varphi_A(t) & = & \varphi_0 - {G_* m_B \alpha_B\over r_{AB}c^2}
+O\left({1\over c^4}\right)\ ,
\label{eq4.8b}\end{eqnarray}
\end{mathletters}
and the standard relations given by Newtonian orbital dynamics (with
effective Newtonian constant $G_{AB} = G_*(1+\alpha_A\alpha_B)$), we
find a usual ``Einstein'' contribution, $\Delta_E = \gamma \sin u$,
to the timing formula \cite{DD86,DT92}. In $\Delta_E$, $u$
denotes the function of $T_A$ defined by solving $u-e \sin u =
2\pi[(T_A-T_0)/P_b - {1\over 2}\dot P_b((T_A-T_0)/P_b)^2]$,
and\footnote{The notation $\gamma^{\rm th}(m_A, m_B)$ in
Eq.~(\ref{eq4.9}) refers to the theoretical prediction, within
tensor--scalar models, giving the phenomenological timing parameter
$\gamma$ as a function of the masses. See below.}
\begin{eqnarray}
\gamma &\equiv& \gamma^{\rm th}(m_A, m_B)
\nonumber\\
&=& {e\over n}\,{X_B\over
1+\alpha_A\alpha_B}
\left({G_{AB}(m_A+m_B) n\over c^3}\right)^{2/3}
\nonumber\\
&& \times \left[X_B(1+\alpha_A \alpha_B)+1+K_A^B\right]\ .
\label{eq4.9}\end{eqnarray}
The timing parameter $\gamma$ should not be confused with the
Eddington parameter $\gamma_{\rm Edd}$. Here $e$ is the orbital
eccentricity, $n \equiv 2\pi/P_b$ the orbital circular frequency,
$X_B \equiv m_B/(m_A+m_B)$, and the new contribution $K_A^B$ coming
from the variation of $I_A$ under the influence of the companion $B$
is defined by
\begin{equation}
K_A^B \equiv -\alpha_B {\partial\ln I_A\over
\partial\varphi_0}\ .
\label{eq4.10}\end{equation}
Note the dissymmetric roles of the labels $A$ and $B$. It is
important, for applications, to recognize that the dependence of the
correction $K_A^B$ upon the two masses $m_A, m_B$ is factorized (in
the single scalar case that we consider here). Accordingly, it might
be convenient to define the quantity
\begin{equation}
k_A(m_A)\equiv -\partial\ln
I_A/\partial\varphi_0\ ,
\label{eq4.11}\end{equation}
so that $K_A^B(m_A,m_B) = k_A(m_A) \alpha_B(m_B)$.

The reasoning above (based on the use of the Einstein conformal
frame) could be done using the ``physical'' (or Jordan--Fierz)
conformal frame. Indeed, the angular momentum is independent of the
conformal frame (being an {\it action\/} variable). This means
$I_A\Omega_A = \widetilde I_A \widetilde \Omega_A$ so that the pulsar
intrinsic time (which is a conformal invariant, being proportional to
the angle $\phi_A$) can be equivalently written as\footnote{Note in
passing that the pulsar clock ticks neither the Einstein time nor the
Jordan--Fierz one. Indeed, both $I_A$ and $\widetilde I_A$ fluctuate
because of their dependence on $\varphi_A(t)$.} $dT_A = C'
d\tau_A^*/I_A = C' d\widetilde\tau_A/\widetilde I_A$. The calculation
is (as always) slightly more complicated in the Jordan--Fierz frame
and leads to a correction $\widetilde K_A^B$ instead of the
$K_A^B$ in Eq.~(\ref{eq4.9}), given by the sum of two terms:
$\widetilde K_A^B = \alpha_0\alpha_B - \alpha_B \partial\ln
\widetilde I_A/\partial\varphi_0$, which is (as it should) found to
be identically equal to $K_A^B$, Eq.~(\ref{eq4.10}), when using
the link $\widetilde I_A = A(\varphi_A) I_A$.

In previous works \cite{E75,W93,DEF1} one had assumed, as an
approximation, that $\widetilde I_A$ was simply a function of the
local, externally imposed, value of the effective gravitational
``constant'' $\widetilde G(\varphi_A)$. Such an assumption is
meaningful only in a ``quasi-weak-field'' approximation where one
formally considers the compactness of a neutron star as a small
expansion parameter (see Section 8.1 of \cite{DEF1}). This
approximation breaks down precisely when the strong-scalar-field
effects studied here develop ({\it i.e.}, when $|\beta| \gtrsim 4$).
Previous treatments introduced the parameter $\kappa_A \equiv
-\partial\ln\widetilde I_A/\partial\ln \widetilde G_A$. When it is
meaningfully defined, the parameter $\kappa_A$ is linked to the
parameter $k_A\equiv -\partial\ln I_A/\partial \varphi_0$ introduced
above by: $k_A \approx \alpha_0 + 2 \alpha_0
[1+\beta_0/(1+\alpha_0^2)]\kappa_A$. This formula shows that, under
the assumptions of previous approximate treatments, the correction
$K_A^B$ was proportional to the product
$\alpha_0\alpha_B$ between the {\it weak-field\/} scalar coupling
$\alpha_0$ and the possibly strong-field amplified effective coupling
$\alpha_B$. As $\alpha_0$ is observationally strongly constrained,
this led always to small values of $K_A^B$ with nearly negligible
observational effects. By contrast, the exact result (\ref{eq4.10}) is
fully sensitive to strong-scalar-field effects taking place both in
the pulsar and its companion. To illustrate the order of magnitude of
possible deviations from the general relativistic
prediction\footnote{The general relativistic prediction $\gamma^{\rm
GR}(m_A,m_B)$ is obtained from Eq.~(\ref{eq4.9}) by setting
$\alpha_A\alpha_B=0$, $G_{AB} = G$, and $K_A^B = 0$.} for the
timing parameter $\gamma$ in systems made of two neutron stars, we
plot in Fig.~\ref{fig4} the value of $K^A_A$ (corresponding to
the cases where $m_B = m_A$) versus $m_A$ within the model
$A_{\overline 6}(\varphi)$, Eq.~(\ref{eq4.4}), and using the same
assumptions as in Fig.~\ref{fig3} (notably a maximally allowed value
of $\varphi_0$, Eq.~(\ref{eq4.5})). We see on Fig.~\ref{fig4} that
when $m_A \gtrsim 1\, m_\odot$, we get very drastic modifications of
the general relativistic prediction for $\gamma$ (except in a small
neighborhood of $m_A \sim 1.3\, m_\odot$ where $K_A^A$
vanishes). In particular, when $1.1 \leq m_A/m_\odot \leq 1.2$,
$K_A^A$ takes largish {\it negative\/} values which change the
sign of the predicted $\gamma^{\rm th}$~! [The minimum value of
$K_A^A$ in Fig.~\ref{fig4} is reached for $m_A= 1.13\, m_\odot$
and equals $K^{A\,{\rm min}}_A = -3.45$, yielding $\gamma^{\rm
th}_{\rm min}(m_A,m_A) = -1.27\, \gamma^{\rm GR}$.] We computed also
$K^A_A$ for smaller values of the external scalar field
$\varphi_0$ and found (as usual by now) that they cause a sharpening
of the ``resonance'' bump in Fig.~\ref{fig4}. For instance, we found
that $K^{A\,{\rm min}}_A = -6.68$ for $\varphi_0 =
\varphi_0^{\rm max}/10$. Paradoxically, smaller values of the
weak-field coupling $\alpha_0$ predict larger values of the
modification $K^A_A$ to the timing parameter $\gamma$, though
concentrated over a smaller range of mass values. This effects is
even more spectacular for $K_A^B$ when $\overline m_B >
\overline m_A\approx \overline m_{\rm cr}$~: in that case, the
effective coupling $\alpha_B$ tends to a non-vanishing constant as
$\varphi_0 \rightarrow 0$, while $\partial\ln I_A/ \partial\varphi_0$
blows up, so that $|K_A^B|$ can take arbitrarily large values.
For instance, one gets $K_A^{B\,{\rm min}} = -8.20$ for
$\varphi_0 = \varphi_0^{\rm max}$, and $K_A^{B\,{\rm min}} =
-23.82$ for $\varphi_0 = \varphi_0^{\rm max}/10$ [yielding
$\gamma^{\rm th}_{\rm min}(m_A,m_B) \approx -23\, \gamma^{\rm GR}$!].
Finally, varying the value of the curvature parameter $\beta$ in the
models (\ref{eq2.1}) displaces the center of the bump in $K^A_A$,
towards lower (higher) values when $-\beta$ increases (decreases).

\section{Application to binary-pulsar experiments}
We have now all the necessary tools in hand for exploring the impact
of non-perturbative scalar effects on binary-pulsar experiments. In a
future publication \cite{DEFT}, we shall confront in a systematic
manner the predictions of tensor--scalar gravitation theories with
a more complete and updated set of binary-pulsar data. In the present
work, we shall illustrate how binary-pulsar data give us very
significant constraints on the strong-field regime of relativistic
gravity by comparing published data on PSR 1913+16, PSR 1534+12 and
PSR\footnote{Note that the pulsar community now uses an updated
notation in which these pulsars are called PSR B 1913+16, PSR B
1534+12 and PSR B 0655+64, respectively. Here, the label B (for
Besselian) refers to the equatorial coordinate system based on the
1950 equinox [while the letter J (for Julian) refers to the 2000
equinox].} 0655+64 with the predictions of tensor--scalar theories
exhibiting the nonperturbative effects discussed above. In Ref.
\cite{DEFT}, we shall also take into account data on certain nearly
circular binary systems which test the strong equivalence principle
\cite{DS91,A95,W95}. We do not consider them here because they are
less constraining than the systems we study. [Indeed, the ``Stark
effect'' is proportional to the product $\alpha_0
(\alpha_A-\alpha_B)$ and, therefore, is already significantly
constrained by the solar-system limits on $\alpha_0$.]

The case of PSR 1913+16 is the richest in that it involves many
different types of strong-field effects: (i)~modifications of the
first post-Keplerian orbital motion (observable through the
periastron advance $\dot\omega$), (ii)~modification of gravitational
radiation damping (observable through the orbital period decay $\dot
P_b$), and (iii)~sensitivity of the pulsar inertia moment to an
external scalar field (observable through the timing parameter
$\gamma$). As we shall discuss below, the case of PSR 1534+12 very
usefully complements that of 1913+16 in trading the $\dot P_b$
measurement against a measurement of the shape parameter $s$ of the
gravitational delay. The scalar-field effects in $\dot\omega$, $\dot
P_b$ and $s$ have been already worked out with sufficient accuracy in
the literature \cite{DEF1,DT92,W93}, while the scalar-field effects in
$\gamma$ have been discussed above.

\subsection{The PSR 1913+16 experiment}
We recall that, at present, one can phenomenologically extract from
the raw data of the binary pulsar PSR 1913+16 (following the
methodology of \cite{DT92}) {\it three\/} well-measured\footnote{Two
more observables, $r^{\rm obs}$, $s^{\rm obs}$, are measured with low
precision \cite{TWDW}.} observables: $\dot\omega^{\rm obs}$,
$\gamma^{\rm obs}$ and $\dot P_b^{\rm obs}$. Here
$\dot\omega^{\rm obs}$ denotes the secular rate of advance of the
periastron, $\gamma^{\rm obs}$ denotes the observed value of the
timing parameter discussed above, and $\dot P_b^{\rm obs}$ denotes
the secular change of the orbital period. The values we shall take
for these observed parameters are \cite{Taylor}:
\begin{mathletters}
\label{eq5.1}
\begin{eqnarray}
\dot\omega^{\rm obs} & = & 4.226621(11)\ {}^\circ\ {\rm yr}^{-1}\ ,
\label{eq5.1a} \\
\gamma^{\rm obs} & = & 4.295(2)\times 10^{-3}\ {\rm s} \ ,
\label{eq5.1b} \\
\dot P_b^{\rm obs} & = & -2.422(6)\times 10^{-12}\ ,
\label{eq5.1c}\end{eqnarray}
\end{mathletters}
where figures in parentheses represent 1$\sigma$ uncertainties in
the last quoted digits. We shall also need the Keplerian parameters
\begin{mathletters}
\label{eq5.2}
\begin{eqnarray}
P_b & = & 27906.9807804(6)\ {\rm s}\ ,
\label{eq5.2a} \\
e & = & 0.6171308(4)\ .
\label{eq5.2b}\end{eqnarray}
\end{mathletters}

The important point (which is the basis of the parame\-trized
post-Keplerian approach \cite{DT92}) is that the observables
(\ref{eq5.1}) have been extracted from the raw pulsar data {\it
without\/} assuming any specific gravitation theory (at least within
the very wide class of boost-invariant theories). One can then use
the three pieces of data (\ref{eq5.1}) to constrain theories of
gravitation. To do this one must compute, within the theory to be
tested, what are the predictions it makes for $\dot\omega$, $\gamma$
and $\dot P_b$ as {\it functions\/} of the two ({\it a priori\/}
unknown) masses $m_A$, $m_B$. We have written in Eq.~(\ref{eq4.9})
above the theoretical prediction for the timing parameter,
$\gamma^{\rm th}(m_A, m_B)$, within tensor--scalar gravity models.
The theoretical prediction for the periastron advance rate has been
worked out in Refs.~\cite{W93,DT92,DEF1} and reads, with the notation
of the present paper,
\begin{eqnarray}
&& \dot \omega^{\rm th}(m_A,m_B) = {3n\over
1-e^2}\left({G_{AB}(m_A+m_B)n\over c^3}\right)^{2/3}
\nonumber\\
&& \times\left[
{1-{1\over 3}\alpha_A\alpha_B\over 1+\alpha_A\alpha_B}
-{X_A \beta_B \alpha_A^2 + X_B \beta_A \alpha_B^2\over
6(1+\alpha_A\alpha_B)^2}
\right]\ .
\label{eq5.3}\end{eqnarray}
The notation in Eq.~(\ref{eq5.3}) is the same as in Eq.~(\ref{eq4.9}).
We recall that $n \equiv 2\pi/P_b$, $G_{AB} =
G_*(1+\alpha_A\alpha_B)$, $X_A \equiv m_A/(m_A+m_B) \equiv 1-X_B$.
Finally, the theoretical prediction for the (radiation-reaction
driven) orbital period decay has been derived in Ref.~\cite{DEF1}
(with the full needed accuracy and for generic tensor--scalar
theories). It is given as a sum of contributions:
\begin{eqnarray}
&& \dot P_b^{\rm th}(m_A,m_B) =
\dot P^{\rm monopole}_\varphi
+ \dot P^{\rm dipole}_\varphi
+ \dot P^{\rm quadrupole}_\varphi
\nonumber\\
&& + \dot P^{\rm quadrupole}_{g*}
+ \left(\dot P^{\rm gal}\right)^{\rm GR}
+ \delta^{\rm th}\left(\dot P^{\rm gal}\right)\ .
\label{eq5.4}\end{eqnarray}
The first three contributions correspond to energy lost to scalar
waves (of monopolar, dipolar and quadrupolar type, respectively). The
fourth one corresponds to the energy lost to quadrupolar tensor waves
(pure spin-2 field $g^*_{\mu\nu}$). The explicit expressions of these
four terms are given in Eqs.~(6.52) of \cite{DEF1}. The fifth
contribution is the value of the galactic contribution to the
observable $\dot P_b$ computed in \cite{DT91} within the assumption
(true in general relativity) that neutron stars fall like ordinary
bodies in the gravitational field of the Galaxy. Finally, the sixth
and last contribution is the modification of the galactic
contribution due to the fact that, in tensor--scalar gravity, neutron
stars fall differently from weakly self-gravitating bodies
(Eq.~(9.22) of \cite{DEF1}).

As usual, given a specific tensor--scalar theory, a value for the
externally imposed asymptotic boundary condition $\varphi_0$, and a
specific nuclear equation of state, the three equations
$\dot\omega^{\rm th}(m_A,m_B)= \dot\omega^{\rm obs}$,
$\gamma^{\rm th}(m_A,m_B)= \gamma^{\rm obs}$,
$\dot P_b^{\rm th}(m_A,m_B)= \dot P_b^{\rm obs}$,
define three curves (in fact three {\it strips}) in the two dimensional
plane of the masses $(m_A,m_B)$. If the three strips meet in a small
region, the considered tensor--scalar theory is consistent with the
binary-pulsar data. If they do not meet, the considered theory is
inconsistent with the pulsar observations.

Before presenting the results of such a confrontation for scalar models
exhibiting non-perturbative effects, it is instructive, for making a
contrast, to discuss two other cases (besides the general relativistic
one): (i)~the case of the well-known Jordan--Fierz--Brans--Dicke
(JFBD) theory \cite{J+,F56,BD61}, and (ii)~the case of scalar--tensor
models with {\it positive\/} curvature of the coupling function. The
JFBD theory contains only one free parameter
$\alpha_0$ and is defined by the coupling function
\begin{equation}
A_{\rm JFBD}(\varphi) = \exp(\alpha_0 \varphi)\ .
\label{eq5.5}\end{equation}
The scalar coupling strength in this theory is constant:
$\alpha(\varphi) \equiv \partial\ln A/\partial\varphi = \alpha_0$.
As a consequence (see Section II), it cannot exhibit nonperturbative
effects. All deviations from general relativity, be they in weak-field
or strong-field conditions, are proportional to $\alpha_0^2$, and are
uniformly constrained by the solar-system limit (\ref{eq1.6}). On the
other hand, as discussed in \cite{DEF3}, scalar--tensor models
belonging to the quadratic class (\ref{eq2.1}) with $\beta >0$ exhibit
nonperturbative effects of a deamplification type: deviations from
general relativity are exponentially suppressed by strong-field
effects, {\it i.e.}, are proportional to $\alpha_0^2
\exp(-3\beta s_A)$, where $s_A\sim 0.2$ is a measure of the strength
of the self-gravity of the considered neutron star.

As one of the aims of the present work is to delineate the cases where
binary-pulsar data give more stringent constraints than solar system
data on tensor--scalar theories, we shall draw the figures below
(except when otherwise indicated) under the assumption that the
externally imposed $\varphi_0$ always takes the maximum value allowed
by solar-system tests. As said above, the corresponding value of
$\varphi_0$ depends on the theory considered and is determined from
combining the two inequalities (\ref{eq1.6}) and (\ref{eq4.6}).
Fig.~\ref{fig5} exhibits the curves defined by the observables
$\dot\omega$, $\gamma$ and $\dot P_b$ in PSR 1913+16 when interpreted
in the framework of three different theories: general relativity,
Jordan--Fierz--Brans--Dicke and the quadratic model (\ref{eq2.1})
with $\beta = +6$. This plot illustrates the fact that binary pulsars
involving nearly equal mass\footnote{As emphasized by Eardley
\cite{E75}, the situation is different for unequal mass systems
thanks to the presence of scalar dipole radiation $\propto
(\alpha_A-\alpha_B)^2$. This shows up in Fig.~\ref{fig5} as a strong
distortion of the $\dot P_b$ curve away from the diagonal $m_A =
m_B$. See below our study of the unequal mass system PSR 0655+64.}
neutron stars do not constrain more severely than solar-system data
theories with logarithmic coupling functions $\ln A(\varphi)$ which
are either linear (or nearly linear) in $\varphi$ \cite{WZ} or have a
{\it positive\/} curvature ({\it i.e.}, convex functions $\ln
A(\varphi)$). Note that the $\dot P_b$ curve of a quadratic model
with positive curvature lies between the general relativistic and
the JFBD ones. The fact that in Fig.~\ref{fig5} the $\dot
P_b^{\beta = +6}$ curve almost overlaps with the $\dot\omega$ curve
is a numerical coincidence caused by our choices for $\beta$ and
$\varphi_0$.

By contrast, tensor--scalar theories involving sufficiently {\it
concave\/} functions $\ln A(\varphi)$, {\it i.e.}, models
(\ref{eq2.1}) with $\beta$ sufficiently negative, show a very
different behavior when confronted to pulsar data. This is
illustrated in the remaining figures. The left panel of
Fig.~\ref{fig6} shows that although the quadratic model
$A_{\overline{4.5}}(\varphi)$ ({\it i.e.}, $\beta = -4.5$ in
Eq.~(\ref{eq2.1})) can pass all present solar-system tests, it fails
the $(\dot\omega\mbox{-}\gamma\mbox{-}\dot P_b)_{1913+16}$ test. For
such a model, pulsar observations constrain more strongly the theory
than weak-field tests. This is further illustrated in the right panel
of this figure, which shows that one needs a smaller value of
$\varphi_0$, {\it i.e.}, a smaller value of $\alpha_0 =
\beta\varphi_0$ than the maximal one allowed by solar data. We did
not make a exhaustive numerical search but it seems that a value
$\alpha_0^{\rm PSR} \sim {1\over 30} \alpha_{0{\rm solar}}^{\rm max}$
is the correct order of magnitude that pulsar data can tolerate.
Note that, in terms of the basic Eddington parameters
$\overline\gamma \equiv \gamma_{\rm Edd} - 1 \approx -2 \alpha_0^2$,
$\overline \beta \equiv \beta_{\rm Edd} - 1 \approx {1\over
2}\beta_0\alpha_0^2$, this means that binary-pulsar data are one
thousand times more constraining than solar-system tests,
constraining $\overline\gamma$ and $\overline\beta$ (for the
considered model $A_{\overline{4.5}}(\varphi)$) below the $10^{-6}$
level.

In the above case ($\beta = -4.5$), the fact that the maximal
weak-field-allowed value of $\varphi_0$ was forbidden, while a
30 times smaller one was allowed, was due to the presence of
significant nonperturbative scalar effects for $\varphi_0 =
\varphi_{0{\rm solar}}^{\rm max}$ which tended to zero when
$\varphi_0 \rightarrow 0$. This smooth disappearance of
nonperturbative effects when $\varphi_0\rightarrow0$ was connected to
the fact that the critical value of the mass $\overline m_{\rm cr}$
(above which spontaneous scalarization occurs in zero-external-field
conditions) is slightly larger that the actual mass of the stars
[$\overline m_{cr}(\beta = -4.5)\approx 1.84\, m_\odot$ while
$\overline m_A^{\rm best\, fit}\approx \overline m_B^{\rm best\,
fit}\approx 1.5\, m_\odot$].

A quite different situation arises for values of $-\beta$ large
enough to make $\overline m_{\rm cr}$ smaller than the masses of the
stars. This case is illustrated in Fig.~\ref{fig7} where one
confronts pulsar data with the model $A_{\overline 6}$,
Eq.~(\ref{eq4.4}). [In this model $\overline m_{\rm cr}(\beta = -6)
\approx 1.24\, m_\odot$.] This figure shows that the model
$A_{\overline 6}$ fails very badly the
$(\dot\omega\mbox{-}\gamma\mbox{-}\dot P_b)_{1913+16}$ test, while
it can pass all present solar-system tests. Especially remarkable is
the wild behavior of the $\gamma$ curve, which is due to the large
values of the $K_A^B$ deviation discussed in the previous
section (see Fig.~\ref{fig4} there). In that case, this disagreement
between theory and experiment is {\it not\/} alleviated by
considering smaller values of $\varphi_0$, as illustrated on the
right panel of Fig.~\ref{fig7}. From our numerical results, we find
that the $(\dot\omega\mbox{-}\gamma\mbox{-}\dot P_b)_{1913+16}$ test
can be passed, if at all, only for extremely fine-tuned values of the
masses in close neighborhoods of the critical values $\overline
m_A\approx\overline m_B\approx\overline m_{\rm cr}$. Barring any
fine-tuned coincidence, we conclude that the tensor--scalar theory
defined by $A_{\overline 6}(\varphi)$ is incompatible with pulsar
data {\it whatever be the value of\/}
$\varphi_0$, even infinitely smaller than $\varphi_{0{\rm
solar}}^{\rm max}$. This remarkable conclusion proves explicitly that
pulsar data are {\it qualitatively\/} different from solar-system
data in their probing power of relativistic gravity. The theory
defined by $A_{\overline 6}(\varphi)$ could always be made compatible
with solar-system tests of any precision\footnote{For the argument's
sake we assume here that general relativity is {\it the\/} correct
theory of gravity.}, while it is already falsified by existing pulsar
observations. As the critical mass $\overline m_{\rm cr}$ decreases
when $-\beta$ increases, the confrontation theory/pulsar-experiments
can only get worse when $-\beta > 6$. Furthermore, our (partial)
numerical exploration of the range $-6 < \beta < -4.5$ shows that
values of $\beta$ smaller than $-5$ are already excluded. [This will
be illustrated by an exclusion plot discussed below.] We conclude
that present PSR 1913+16 data already exclude all ``quadratic''
models (\ref{eq2.1}) with $\beta < -5$.

\subsection{The PSR 1534+12 experiment}
One conceivable deficiency in the above argument is the possible
presence of a cosmological variation of $\varphi_0$. Indeed, a
nonzero value of $\dot \varphi_0 \equiv d\varphi_0/dt_0$ entails a
secular variation of the strength of scalar gravity and produces an
additional contribution in $\dot P_b^{\rm th}$. From the observed
$\dot P_b^{\rm obs}$, one cannot decorrelate $\dot
\varphi_0$-effects from scalar modifications to radiation damping.
One way to decorrelate $\dot\varphi_0$-effects is to consider pulsar
experiments which are insensitive to $\dot\varphi_0$. Such is the
case of the measurements of the three observables $\dot\omega^{\rm
obs}$, $\gamma^{\rm obs}$ and $s^{\rm obs}$ in PSR 1534+12. Here
$s^{\rm obs}$ denotes a phenomenological parameter measuring the
shape of the gravitational time delay \cite{DD86,DT92}. The values
we shall take for these three\footnote{We do not consider here the
other observables $r^{\rm obs}$ and $\dot P_b^{\rm obs}$ which are
measured with low fractional precision.} observable parameters are
\cite{A95}
\begin{mathletters}
\label{eq5.6}
\begin{eqnarray}
\dot\omega^{\rm obs} & = & 1.75573(4)\ {}^\circ\ {\rm yr}^{-1}
\label{eq5.6a} \\
\gamma^{\rm obs} & = & 2.081(16) \times 10^{-3} \ ,
\label{eq5.6b} \\
s^{\rm obs} & = & 0.981(8) \ .
\label{eq5.6c}\end{eqnarray}
\end{mathletters}
We shall also need the Keplerian observables
\begin{mathletters}
\label{eq5.7}
\begin{eqnarray}
P_b & = & 36351.70267(2)\ {\rm s}\ ,
\label{eq5.7a} \\
e & = & 0.2736771(4)\ ,
\label{eq5.7b} \\
x & = & 3.729458(2)\ {\rm s} \ .
\label{eq5.7c}\end{eqnarray}
\end{mathletters}
The theoretical predictions for $\dot\omega^{\rm th}$ and
$\gamma^{\rm th}$ have been written above. That for $s^{\rm th}$
reads \cite{DT92}
\begin{equation}
s = {n x \over X_B}[G_{AB} (m_A+m_B) n /c^3]^{-1/3}\ ,
\label{eq5.8}\end{equation}
where we have used the same notation as in Eqs.~(\ref{eq4.9}) and
(\ref{eq5.3}) above, and where $x = a_1 s/c$ is the projection of the
semi-major axis ($a_1$) of the pulsar orbit on the line of sight (in
light-seconds).

We have plotted the three curves defined by this
$(\dot\omega\mbox{-}\gamma\mbox{-}s)_{1534+12}$ test for various
values of $\beta$ and $\varphi_0$. For instance, we exhibit the case
$\beta = -6$, $\varphi_0 = \varphi_{0{\rm solar}}^{\rm max}$ in Fig.
\ref{fig8} (together with the case of general relativity). From our
(partial) numerical study, we conclude that the quadratic models
$A_{\beta}$ fail the $(\dot\omega\mbox{-}\gamma\mbox{-}s)_{1534+12}$
test when $\beta < -5.5$. The corresponding exclusion plot is very
similar to that defined by PSR 1913+16 (see below).

\subsection{The PSR 0655+64 experiment}
The binary pulsar PSR 0655+64 is composed of a neutron star of
mass $\approx 1.4\, m_\odot$, and a white dwarf companion of mass
$\approx 0.8\, m_\odot$. They move around each other on a nearly
circular orbit in a period of about one day. In tensor--scalar
gravity, such a dissymmetrical system is a powerful emitter of dipolar
scalar waves, especially in presence of nonperturbative scalar
effects. The theoretical prediction \cite{DEF1} for the corresponding
orbital period decay is dominated by the $O(v^3/c^3)$ dipole
contribution in Eq.~(\ref{eq5.4}) above:
\begin{eqnarray}
&&\dot P_b^{\rm th}(m_A,m_B) \approx
\dot P_\varphi^{\rm dipole}
= -{2\pi G_* m_A m_B n \over (m_A+m_B)c^3}
\nonumber\\
&&\times
{1+e^2/2\over (1-e^2)^{5/2}}\,\left(\alpha_A-\alpha_B\right)^2
+O\left({v^5\over c^5}\right)\ .
\label{eq5.9}\end{eqnarray}
The fact that the observed value of $\dot P_b$ in PSR 0655+64 is
very small (and, in fact, consistent with zero) constrains very much
the magnitude of the effective coupling strength $\alpha_A$, and
therefore the possibility of nonperturbative effects. The
experimental data we need for our analysis are taken from
Ref.~\cite{A95}:
\begin{mathletters}
\label{eq5.10}
\begin{eqnarray}
P_b & = & 88877.06194(4)\ {\rm s}\ ,
\label{eq5.10a} \\
e & < & 3\times 10^{-5}\ ,
\label{eq5.10b} \\
x & \equiv & (a_1\sin i)/c = 4.12560(2)\ {\rm s}\ ,
\label{eq5.10c} \\
\dot P_b & = & (1\pm 4)\times 10^{-13}\ .
\label{eq5.10d}\end{eqnarray}
\end{mathletters}
The masses of the pulsar and its companion are not known
independently. From the observed mass function, the {\it a priori\/}
statistics of the inclination angle $i$, and the observed small
statistical spread of neutron star masses around $1.35\, m_\odot$, one
can deduce a range of probable values for the pair $(m_A,m_B)$~:
Essentially, one is limited to a sub-region of the rectangle $m_A =
(1.35\pm 0.05)m_\odot$, $m_B = (0.8\pm 0.1)m_\odot$ in the mass
plane\footnote{We use here the fact that the scalar
modifications to the link between the observed mass function
$n^2(a_1\sin i)^3$ and the Einstein masses $m_A$, $m_B$ due to
the factor $G_{AB}/G = 1+\alpha_A\alpha_B$ are small because
$\alpha_B\approx\alpha_0$ for the white dwarf companion.}
\cite{A95}. In our calculations, we will choose the mass pair which
gives the most conservative bounds on tensor--scalar gravity, namely
$m_A = 1.30\, m_\odot$, $m_B = 0.7\, m_\odot$.

Finally, using the fact that the self-gravity of the white dwarf
companion is negligible compared to that of the pulsar (so that
$\alpha_B\approx\alpha_0$), we get from Eqs.~(\ref{eq5.9}) and
(\ref{eq5.10}) the 1-$\sigma$ level constraint
\begin{equation}
\bigl(\alpha_A(m_A)-\alpha_0\bigr)^2 < 3\times 10^{-4}\ .
\label{eq5.11}\end{equation}

\subsection{Exclusion plots within a generic two-dimensional plane
of tensor--scalar theories}
It is instructive to contrast the pulsar constraints on
tensor--scalar gravity with the constraints obtained from
solar-system experiments. We can use the class of quadratic models
(\ref{eq2.1}) as a generic description of the shape of the coupling
function around the current cosmological value of $\varphi$. In other
words, we can parametrize an interesting class of tensor--scalar
models by two parameters\footnote{This two-parameter class of
models is representative of the large class of coupling functions
$A(\varphi)$ which admit a local minimum and contain no large
dimensionless parameters [{\it i.e.}, we assume that higher
derivatives $\beta'_0\equiv
\partial\beta(\varphi_0)/\partial\varphi_0$, $\beta''_0\equiv
\partial\beta'(\varphi_0)/\partial\varphi_0$ are of order
unity].}: say, $\alpha_0\equiv \alpha(\varphi_0)$ and
$\beta_0\equiv\partial\alpha(\varphi_0)/\partial\varphi_0$. [In
quadratic models, $\alpha_0 = \beta\varphi_0$, and $\beta_0 = \beta$
is field independent.] We can then interpret all experimental data
(solar-system and pulsar ones) as constraints in the
two-dimensional theory plane $(\alpha_0,\beta_0)$. For instance,
neglecting the correlations in the measurements of the two Eddington
parameters $\gamma_{\rm Edd}$ and $\beta_{\rm Edd}$, solar-system
data rule out the regions of the $(\alpha_0,\beta_0)$ plane where
the inequalities $|\gamma_{\rm Edd}-1|<2\times 10^{-3}$ ({\it i.e.},
$\alpha_0^2<10^{-3}$) \cite{gamma}, and $|\beta_{\rm Edd}-1|<6\times
10^{-4}$ ({\it i.e.}, $|\beta_0|\alpha_0^2<1.2\times10^{-3}$)
\cite{LLR} are not satisfied. The corresponding exclusion plot is
represented in Fig.~\ref{fig9}. On the same plot, we can represent
the constraints brought by pulsar data on tensor--scalar models. The
PSRs 1913+16 and 1534+12 data give constraints which are numerically
similar. Taken together, they exclude the region to the left of the
wavy line indicated on Fig.~\ref{fig9}. This line is approximate and
was obtained by interpolating between a few values of $\alpha_0$ and
$\beta_0$. We leave to future work a detailed study of the precise
region of the $(\alpha_0,\beta_0)$-plane excluded by these pulsar
data.

As for the PSR 0655+64 data, they define through Eq.~(\ref{eq5.11})
(with the conservative value $m_A = 1.30\, m_\odot$) another limit on
tensor--scalar models. We have numerically computed the region of
the $(\alpha_0,\beta_0)$-plane defined by the inequality
(\ref{eq5.11}). The corresponding allowed region is contained within
the two dashed lines labeled 0655+64 on Fig.~\ref{fig9}. The region
simultaneously allowed by all the tests is shaded.

To prevent any confusion, let us note that the limit on the 2PN
parameter $\zeta\equiv \beta_0^2\alpha_0^2/(1+\alpha_0^2)^3$
obtained in a recent simplified {\it analytical\/} study of combined
pulsar data \cite{DEF2PN}, is valid only for $|\beta|\lesssim 1$.
Indeed, the approximate treatment of \cite{DEF2PN} assumed the
absence of any nonperturbative effect, {\it i.e.}, an absolute value
of $\beta_0$ appreciably smaller than 4. The PSR 0655+64 constraint
studied here should merge with the limit $\beta_0^2\alpha_0^2<
4\times 10^{-3}$ derived in \cite{DEF2PN} when $|\beta_0|\lesssim
1$. [We use here the inequality (5.24c) of Ref. \cite{DEF2PN}
together with the theoretical constraint that $\beta_0^2\alpha_0^2$
be positive.] This merging occurs anyway in a region which is
already excluded by solar-system data.

\section{Conclusions\label{SEC6}}
Before summarizing the main results of the present work, we wish to
indicate briefly the frameworks within which our findings might be
physically relevant. First, let us note that (barring any unnatural
fine tuning) they are not relevant in the case of a strictly massless
scalar field $\varphi$, or at least when its mass $m_\varphi \ll
\hbar H_0/c^2 \sim h_{100} \times 2.13\times 10^{-33}\ {\rm eV}$,
where $H_0 = h_{100}\times 100\ {\rm km}.{\rm s}^{-1}/{\rm Mpc}$
denotes Hubble's constant. Indeed, in such a case it was found, both
in traditional equivalence-principle-respecting tensor--scalar
theories \cite{DN93} and in string-inspired models with a massless
dilaton or modulus \cite{DP94}, that the cosmological evolution
naturally drives the vacuum expectation value of $\varphi$ toward
{\it minima\/} of $\ln A(\varphi)$. As the latter vacuum expectation
value coincides (modulo small {\it fractional\/} corrections due to
spatial fluctuations of the gravitational potential \cite{DEF2PN})
with our externally-imposed $\varphi_0$, a natural prediction of
these massless models is that $\alpha_0= \partial\ln
A(\varphi_0)/\partial\varphi_0$ is small and that $\beta_0 =
\partial^2\ln A(\varphi_0)/\partial\varphi^2_0$ is {\it positive}.
The former result concerning $\alpha_0$ is in agreement with
observational data, but the latter one concerning $\beta_0$ gives the
wrong sign for spontaneous scalarization
effects\footnote{Note, however, that a coupling function of the type
$\ln A(\varphi)= +\epsilon^2\varphi^2 - \lambda^2\varphi^4$, with a
sufficiently small $\epsilon$ and a sufficiently large $\lambda$,
would reconcile a (very localized) minimum at $\varphi=0$ with a
mainly concave coupling function leading to nonperturbative effects.
We do not wish to consider here such fine-tuned cases.}.

On the other hand, our results are relevant to the wide class of
models comprising scalar fields of range\footnote{The lower bound
comes from the requirement that $\varphi$ be effectively massless on
the scale of a typical binary pulsar. If it is violated one must
correct our formulas by Yukawa exponential factors.} $10^6\ {\rm
km}\lesssim \lambda = \hbar/m_\varphi c \lesssim c H_0^{-1}$. Assuming
that the endemic ``Polonyi problem'' (too much energy stored in the
cosmological oscillations of $\varphi_0(t)$) \cite{Polonyi1} of such
models is somehow solved \cite{Polonyi2} or fine-tuned to provide
$\Omega = \rho_{\rm tot}/\rho_{\it cr}=1$ \cite{Frieman}, the
condition for these models to be naturally compatible with
solar-system constraints is simply that the location $\varphi_0$ of
the minimum of the $\varphi$ potential, $V(\varphi) = (c^4/8\pi G_*)
m_\varphi^2 (\varphi-\varphi_0)^2$, coincide (or nearly coincide)
with an {\it extremum\/} of the coupling function $A(\varphi)$. Such
a coincidence would, for instance, naturally follow from a discrete
symmetry about $\varphi_0$, say a reflection symmetry
$(\varphi-\varphi_0) \rightarrow -(\varphi-\varphi_0)$ \cite{DP94}.
Under such a condition, the present value of the weak-field scalar
coupling $\alpha_0 = \partial\ln A(\varphi_0)/\partial\varphi_0$ would
be naturally extremely small, and the sign of the curvature $\beta_0=
\partial^2\ln A(\varphi_0)/\partial\varphi_0^2$ could be expected to
be negative with {\it a priori\/} 50~\% probability.

As a simple example of such models, we can consider a finite-range
scalar field coupled only to the gravitational sector through a
multiplicative coupling to the scalar curvature, say
\begin{eqnarray}
S &=& {c^4\over 16\pi G_*} \int {d^4 x\over c}\, \widetilde
g^{1/2}\Bigl( -Z(\Phi)\widetilde g^{\mu\nu} \partial_\mu\Phi
\partial_\nu\Phi - U(\Phi)
\nonumber\\
&&+ F(\Phi)\widetilde R\Bigr)
+ S_m\left[\psi_m;\widetilde g_{\mu\nu}\right]\ ,
\label{eq6.1}\end{eqnarray}
where we have introduced the possibility of an arbitrary
field-dependent normalization of the kinetic term of $\Phi$. One
transforms (\ref{eq6.1}) into our canonical form (\ref{eq1.1})
[completed by a $\varphi$-potential term $V(\varphi) = (c^4/16\pi
G_*)F^{-2}(\Phi) U(\Phi)$] by defining
\begin{mathletters}
\label{eq6.2}
\begin{eqnarray}
g^*_{\mu\nu} & = & F(\Phi)\widetilde g_{\mu\nu}\ ,
\label{eq6.2a} \\
\varphi & = &
\int d\Phi\left[
{3\over 4}\,
{F'^2(\Phi)\over F^2(\Phi)}
+ {1\over 2}\,
{Z(\Phi)\over F(\Phi)}
\right]^{1/2}\ .
\label{eq6.2b}\end{eqnarray}
\end{mathletters}
This corresponds to a coupling function
\begin{equation}
A(\varphi) = F^{-1/2}(\Phi)\ .
\label{eq6.3}
\end{equation}
The simplest example of such models is a massive scalar having a
nonminimal dimensionless coupling to curvature:
\begin{eqnarray}
S &=& {c^4\over 16\pi G_*} \int {d^4 x\over c}\, \widetilde
g^{1/2}\bigl( -\widetilde g^{\mu\nu} \partial_\mu\Phi
\partial_\nu\Phi - m_\Phi^2 \Phi^2
\nonumber\\
&& + \xi\widetilde R\Phi^2 + \widetilde R\, \bigr)
+ S_m\left[\psi_m;\widetilde g_{\mu\nu}\right]\ .
\label{eq6.4}\end{eqnarray}
This corresponds to $Z(\Phi) = 1$, $U(\Phi) = m_\Phi^2 \Phi^2$, and
$F(\Phi) = 1+\xi\Phi^2$. In that case, one can integrate
Eq.~(\ref{eq6.2b}) explicitly. Assuming for instance
that $\xi(1+6\xi) > 0$ and introducing the notation $\chi \equiv
\sqrt{\xi(1+6\xi)}$, one gets
\begin{eqnarray}
&& 2\sqrt{2}\,(\varphi-\varphi_0) =
{\chi\over\xi}\,
\ln\left[
1 + 2\chi\Phi
\left(
\sqrt{1+\chi^2\Phi^2}
+\chi\Phi\right)
\right]
\nonumber \\
&&+{\sqrt{6}}\,
\ln\left[
1 - 2\sqrt{6}\,\xi\Phi\,
{\sqrt{1+\chi^2\Phi^2}
-\sqrt{6}\,\xi\Phi
\over
1+\xi\Phi^2}
\right]\ .
\label{eq6.5}\end{eqnarray}
Note that $(\varphi-\varphi_0) = \Phi/\sqrt{2} + O(\Phi^3)$ when
$\Phi\rightarrow 0$. From Eqs.~(\ref{eq6.2})--(\ref{eq6.3}),
one gets easily the following parametric representations
\begin{mathletters}
\label{eq6.6}
\begin{eqnarray}
A(\varphi) &=& \left(
1+ \xi\Phi^2
\right)^{-1/2}\ ,
\label{eq6.6a}\\
\alpha(\varphi) &=&
{\partial \ln A(\varphi)\over \partial \varphi}
\nonumber\\
&=&
-\sqrt{2}\,\xi\Phi\left[
1 + \xi(1+6\xi)\Phi^2
\right]^{-1/2}\ ,
\label{eq6.6b}\\
\beta(\varphi) &=&
{\partial^2 \ln A(\varphi)\over \partial \varphi^2} =
-2\xi\left(
1+\xi\Phi^2
\right)
\nonumber\\
&& \times
\left[
1+\xi(1+6\xi)\Phi^2
\right]^{-2}\ .
\label{eq6.6c}
\end{eqnarray}
\end{mathletters}
In particular the value of the curvature parameter around $\varphi_0$
reads
\begin{equation}
\beta_0 = \beta(\varphi_0)
 = \left.{\partial^2 \ln A(\varphi)\over
\partial\varphi^2}\right|_{\Phi=0}= -2\xi\ .
\label{eq6.7}\end{equation}
Therefore, positive values of $\xi$ [which, in the formulation
(\ref{eq6.4}), seem preferred because unable to cause a change of sign
of the coefficient $(1+\xi\Phi^2)$ of the kinetic term for
$\widetilde g_{\mu\nu}$] correspond to the ``interesting'' case where
spontaneous scalarization effects can occur\footnote{In our
conventions, the so-called ``conformal coupling'' corresponds to $\xi
= -{1\over 6}$, and to a coupling function $A(\varphi) =
\cosh(\varphi/\sqrt{3})$. Note, however, that only the action
$S_{\rm conf}\equiv \int d^4 x\, \widetilde g^{\,1/2}[-\widetilde
g^{\mu\nu} \partial_\mu\Phi \partial_\nu\Phi-{1\over 6}\widetilde
R\Phi^2]$ (without bare Einstein term) exhibits a conformal
invariance\cite{Deser}. The action $S_{\rm conf}$ involves a spin-2
but no spin-0 excitation.}.

\begin{figure}
\begin{center}\leavevmode\epsfbox{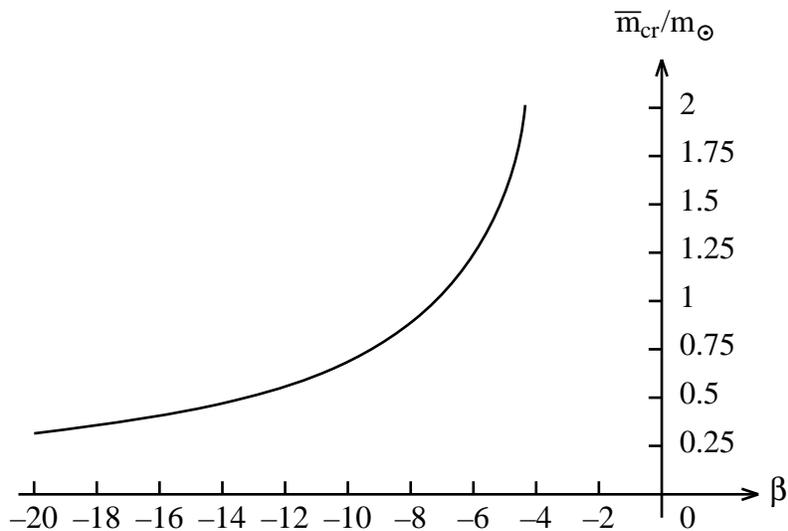}\end{center}\vskip 1pc
\caption{Critical baryonic mass $\overline m_{\rm cr}$
versus the curvature parameter $\beta$ within the quadratic models
$A(\varphi) = \exp({1\over 2}\beta \varphi^2)$.}
\label{fig2}
\end{figure}

\begin{figure}
\begin{center}\leavevmode\epsfbox{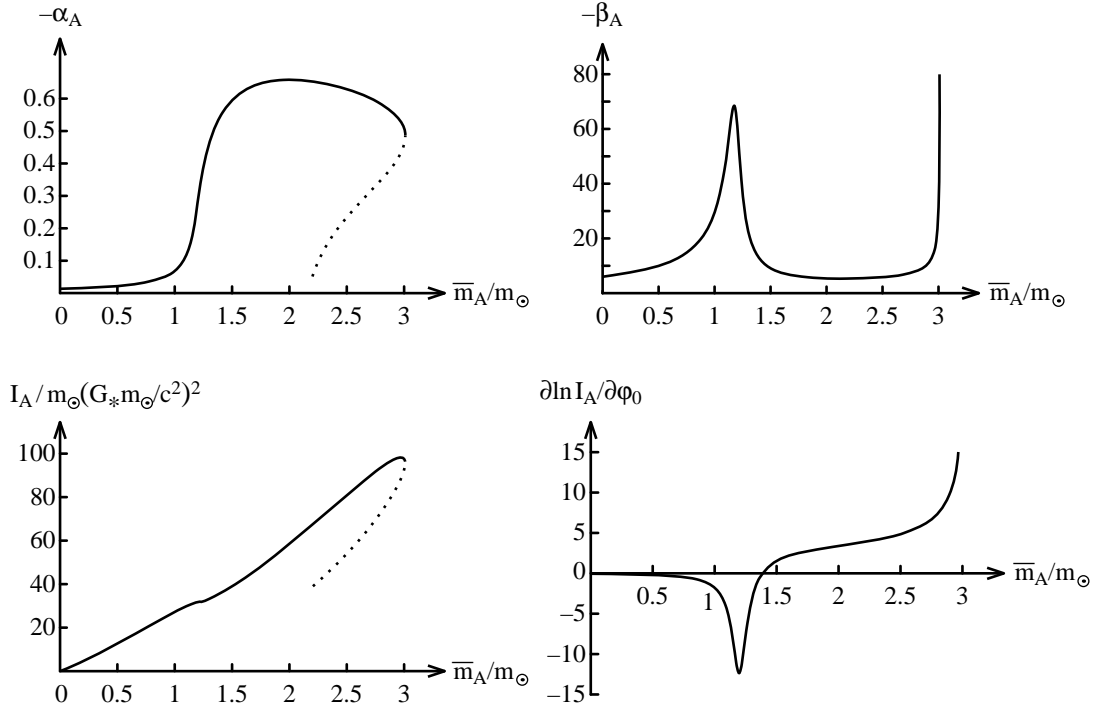}\end{center}\vskip 1pc
\caption{Dependence upon the baryonic mass $\overline m_A$ of the
coupling parameters $\alpha_A$, $\beta_A$, the inertia moment
$I_A$, and its derivative $\partial\ln I_A/\partial\varphi_0$.
These plots correspond to the model $A(\varphi) = \exp(-3
\varphi^2)$ and the maximum value of $\varphi_0$ allowed by
solar-system experiments. As in Fig.~\protect\ref{fig1}, the dotted
lines correspond to unstable configurations of the star.}
\label{fig3}
\end{figure}

\begin{figure}
\begin{center}\leavevmode\epsfxsize=350pt\epsfbox{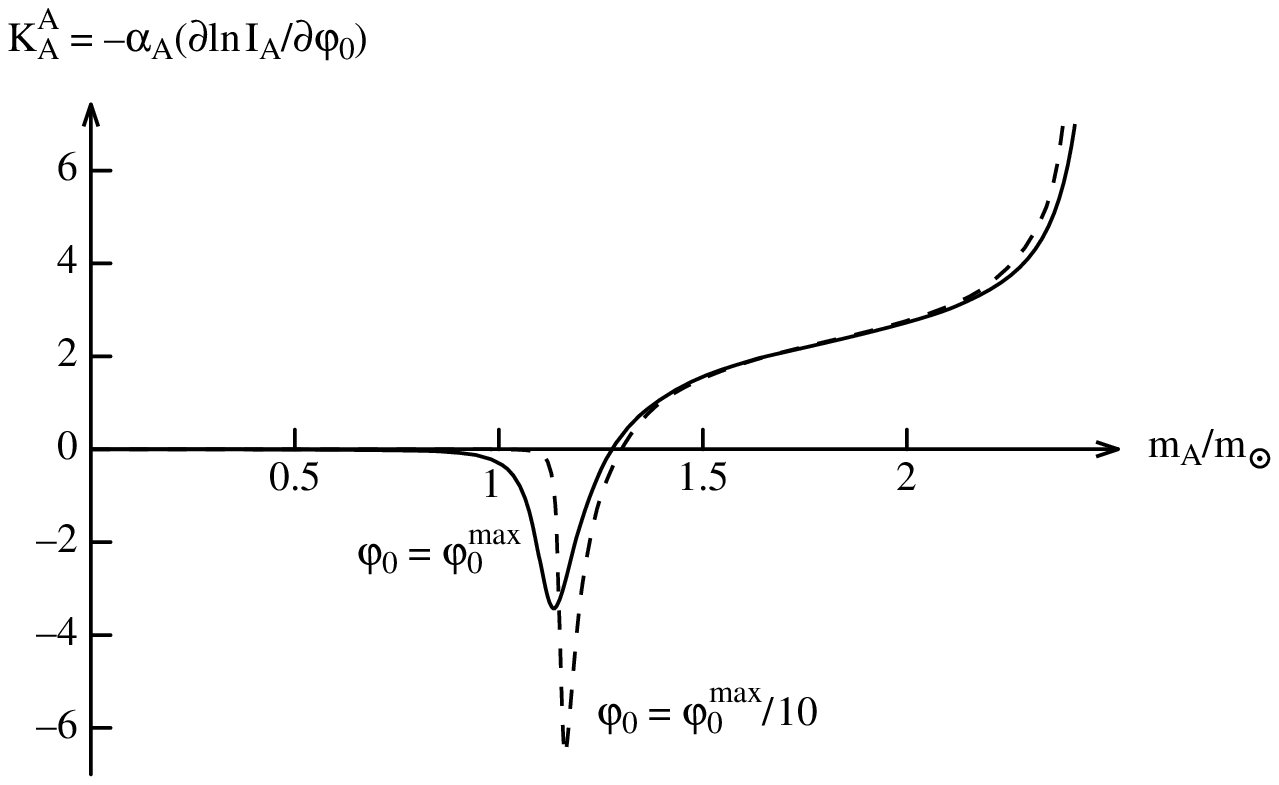}
\end{center}\vskip 1pc
\caption{Parameter $K^A_A = -\alpha_A(\partial \ln
I_A/\partial \varphi_0)$ versus the Einstein inertial mass $m_A$,
within the model $A(\varphi) = \exp(-3 \varphi^2)$. The solid
line corresponds to the maximum value of $\varphi_0$ allowed by
solar-system experiments, and the dashed line to a ten-fold
smaller value of $\varphi_0$ ({\it i.e}, a 100 times smaller
value of the Eddington parameter $\gamma_{\rm Edd}-1$).}
\label{fig4}
\end{figure}

\begin{figure}
\begin{center}\leavevmode\epsfxsize=280pt\epsfbox{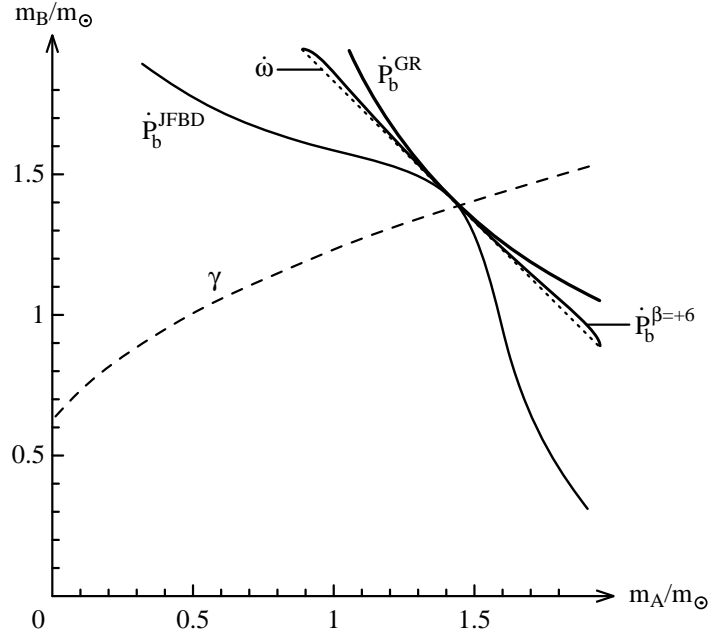}
\end{center}\vskip 1pc
\caption{The $(\dot\omega\mbox{-}\gamma\mbox{-}\dot P_b)_{1913+16}$
test for general relativity (GR), the Jordan--Fierz--Brans--Dicke
theory (JFBD), and the quadratic model $A(\varphi) = \exp(+3
\varphi^2)$ [corresponding to a positive curvature parameter $\beta =
+6$]. The widths of the three $\dot P$ lines correspond to
1-$\sigma$ standard deviations. The $\dot \omega^{\rm th} = \dot
\omega^{\rm exp}$ and $\gamma^{\rm th}=\gamma^{\rm exp}$ lines are
wider than 1-$\sigma$ errors, and cannot be distinguished for the
three theories.}
\label{fig5}
\end{figure}

\begin{figure}
\begin{center}\leavevmode\epsfxsize=468pt\epsfbox{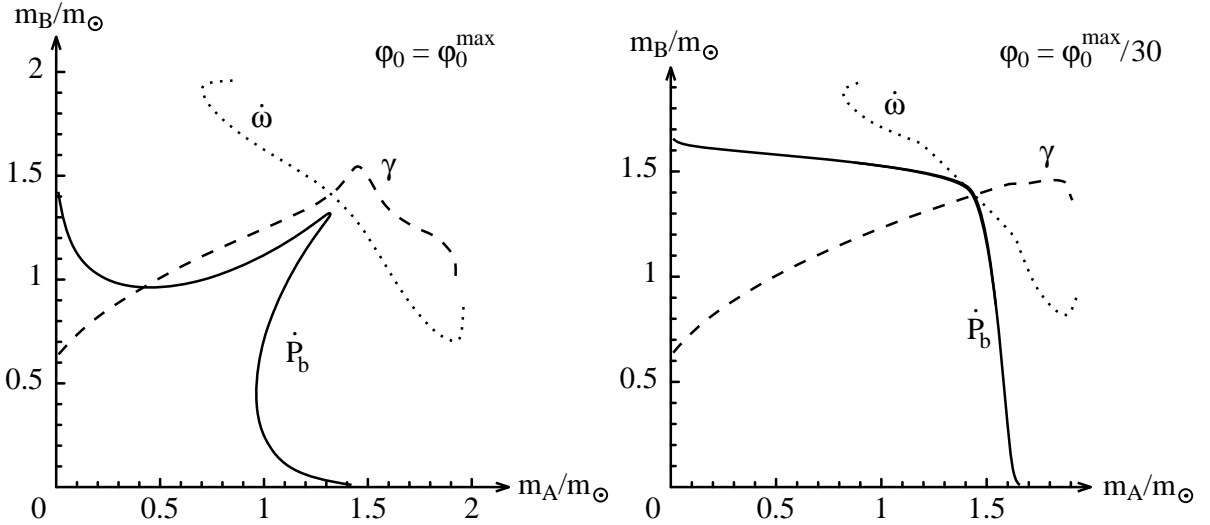}
\end{center}\vskip 1pc
\caption{The $(\dot\omega\mbox{-}\gamma\mbox{-}\dot P_b)_{1913+16}$
test for the quadratic model $A(\varphi) = \exp({1\over 2}\beta
\varphi^2)$ with $\beta = -4.5$, when $\varphi_0$ takes the
maximum value allowed by solar-system experiments (left panel), and
a 30 times smaller value (right panel). In this Figure and the
following ones, 1-$\sigma$ deviations are smaller than the width of
the lines.}
\label{fig6}
\end{figure}

\begin{figure}
\begin{center}\leavevmode\epsfxsize=468pt\epsfbox{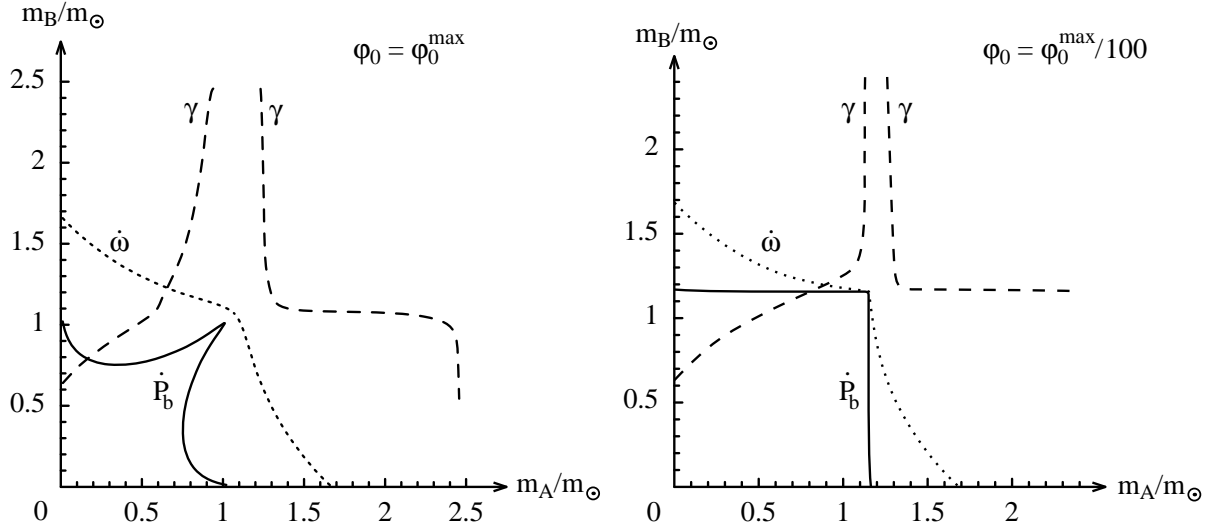}
\end{center}\vskip 1pc
\caption{The $(\dot\omega\mbox{-}\gamma\mbox{-}\dot P_b)_{1913+16}$
test for the quadratic model $A(\varphi) = \exp(-3 \varphi^2)$ [{\it
i.e.}, $\beta = -6$], when $\varphi_0$ takes the maximum value
allowed by solar-system experiments (left panel), and a 100 times
smaller value (right panel).}
\label{fig7}
\end{figure}

\begin{figure}
\begin{center}\leavevmode\epsfxsize=280pt\epsfbox{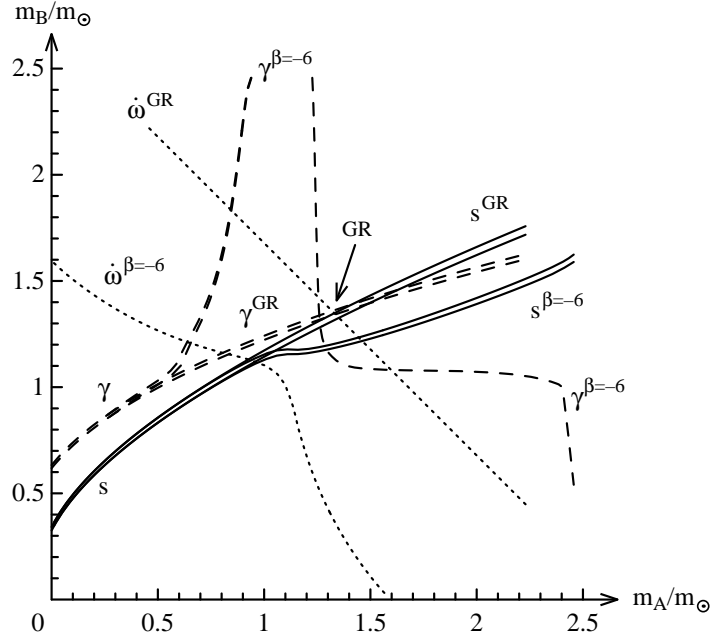}
\end{center}\vskip 1pc
\caption{The $(\dot\omega\mbox{-}\gamma\mbox{-}s)_{1534+12}$
test for general relativity (GR), and for the quadratic model
$A(\varphi) = \exp(-3 \varphi^2)$ [{\it i.e.}, $\beta = -6$] when
$\varphi_0$ takes the maximum value allowed by solar-system
experiments. The widths of the strips correspond to 1-$\sigma$
standard deviations. The arrow indicates the intersection of the
three strips in general relativity. In the model $\beta = -6$, the
three strips do not intersect.}
\label{fig8}
\end{figure}

Let us now summarize the main results of the present work:

$\bullet$~We have clarified the physical meaning of the
nonperturbative strong-gravitational-field effects discovered in
\cite{DEF3} by interpreting them, by analogy with ferromagnetism, as
a phenomenon of ``spontaneous scalarization''. Negative values of the
curvature parameter $\beta_0$ (like negative values of the coupling
between magnetic moments $H_{ij} = g \mbox{\boldmath$\mu$}_i
\mbox{\boldmath$\mu$}_j$) favor the spontaneous creation of a scalar
field when considered in the context of gravitationally compact
objects (neutron stars). The critical baryonic mass for the
spontaneous scalarization transition is $\lesssim 1.5\, m_\odot$ when
$\beta_0\lesssim -5$. See Table~\ref{tab1} and Fig.~\ref{fig2} for
precise values. The presence of some externally-imposed scalar field
background $\varphi_0$ with $\alpha_0 = \partial\ln
A(\varphi_0)/\partial\varphi_0\neq 0$ smoothes the scalarization
transition (analogously to the presence of an external magnetic field
for the ferromagnetic transition).

$\bullet$~The development (through the previous mechanism) of strong
scalar fields in neutron stars leads to very significant deviations
from general relativity. These deviations are measured by some
gravitational form factors $\alpha_A,\beta_A,K_A^B$ which enter
the effects observable in binary-pulsar experiments. In our previous
work \cite{DEF3} we focussed on the effective scalar coupling constant
$\alpha_A$. Here, we gave results for $\beta_A =
\partial\alpha_A/\partial\varphi_0$ (see also \cite{gef93}), and for
$K_A^B = -\alpha_B \partial\ln I_A/\partial\varphi_0$ which
enters the parametrized post-Keplerian timing parameter $\gamma$. To
compute $K_A^B$, we generalized to a tensor--scalar context the
work of Hartle \cite{H67} on the general relativistic inertia moments
$I_A$ of slowly rotating neutron stars. We found that $K_A^B$
could cause very drastic deviations from general relativity in
tensor--scalar theories containing no large dimensionless parameters.
This is achieved without fine-tuning and in theories having only
positive-energy excitations.

\begin{figure}
\begin{center}\leavevmode\epsfbox{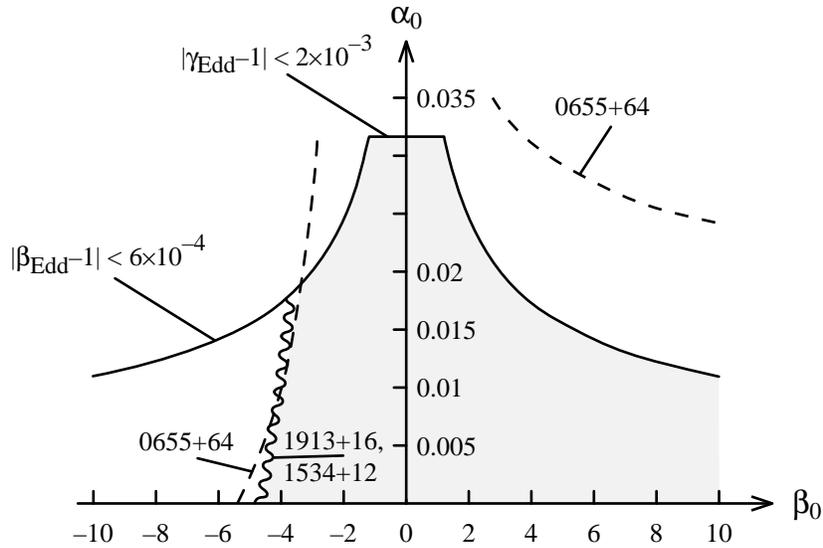}\end{center}\vskip 1pc
\caption{Regions of the $(\alpha_0,\beta_0)$-plane allowed by
solar-system experiments and three binary-pulsar experiments.
In view of the reflection symmetry $\alpha_0\rightarrow -\alpha_0$,
we plot only the upper half plane. The region allowed by solar-system
tests is below the solid line. The PSR 0655+64 test constrains the
values of $\alpha_0$ and $\beta_0$ to be between the two dashed
lines. The region allowed by the PSRs 1913+16 and 1534+12 tests lies
to the right of the (approximate) wavy line. The region
simultaneously allowed by all the tests is shaded.}
\label{fig9}
\end{figure}

$\bullet$~We presented a preliminary investigation of the
confrontation between scalar models exhibiting nonperturbative
effects and actual binary-pulsar experiments. We contrasted the
probing power of pulsar experiments to that of solar-system ones by
working in the two-dimensional $(\alpha_0 \equiv \partial\ln
A(\varphi_0)/\partial\varphi_0 ,\ \beta_0 \equiv \partial^2\ln
A(\varphi_0)/\partial\varphi_0^2)$-plane describing a generic class
of tensor--scalar models. Using published data on PSRs 1913+16,
1534+12 and 0655+64 (and a specific nuclear equation of state), we
found that binary-pulsar experiments exclude a large domain of
theories compatible with solar-system experiments (see the
exclusion plot, Fig.~\ref{fig9}). In particular, they constrain
$\beta_0$ (independently of $\alpha_0$) to
\begin{equation}
\beta_0 > -5\ .
\label{eq6.8}\end{equation}
Interestingly, this bound can be expressed in terms of the well-known
weak-field Eddington parameters
\begin{equation}
{\beta_{\rm Edd}-1\over \gamma_{\rm Edd} -1} < 1.3\ .
\label{eq6.9}\end{equation}
The singular ($0/0$) nature of the ratio on the left-hand side
vividly expresses the fact that such a conclusion could never be
obtained in weak-field experiments (at least until they find a
significant deviation from general relativity). It must be kept in
mind that the inequality (\ref{eq6.9}) is one sided only, and that
$\beta_{\rm Edd} -1$ and $\gamma_{\rm Edd}-1$ must be taken with
their signs.

Let us note that, on the whole, the fact that pulsar data tend to
exclude sufficiently negative values of $\beta_0$ is nicely
compatible with the expectation from cosmological attractor scenarios
\cite{DN93,DP94} that $\varphi_0$ be dynamically driven toward a {\it
minimum\/} of $\ln A(\varphi)$. Though our results have been derived
by assuming a particularly simple coupling function between the
scalar field and matter (``quadratic model'': $\ln A(\varphi) =
{1\over 2} \beta_0 \varphi^2$), we think they hold in general classes
of coupling functions containing no small or large dimensionless
parameters. In a future publication \cite{DEFT}, we shall present a
more systematic confrontation between tensor--scalar theories and
binary-pulsar experiments.

Before ending this paper, we would like to stress some of the
limitations of our treatment that we intend to overcome in future
work: (i)~We considered here only one specific equation of state,
modeled as a simple polytrope; a more complete study should
consider a selection of realistic equations of state. (ii)~In the
present paper, we did not consider the effect of a cosmological
variation of $\varphi_0$. We are aware that a nonzero $\dot
\varphi_0$ of order the Hubble parameter could significantly
modify the interpretation of some of the pulsar tests discussed
above. However, as one disposes of several independent pulsar tests,
some of which do not involve $\dot \varphi_0$-sensitive observables
(such as the $(\dot\omega\mbox{-}\gamma\mbox{-}s)_{1534+12}$
test considered above and several ``Stark'' tests
\cite{DS91,A95,W95}), we think that a combined discussion of all
pulsar tests will lead to limits on $\alpha_0$ and $\beta_0$ similar
to the ones we have obtained here, and, in addition, to significant
limits on $\dot\varphi_0$.

Finally, let us note that the strong scalar field effects discussed
in \cite{DEF3} and the present paper could have a very significant
impact on several aspects of the theory of gravitational radiation
from compact objects (besides the ones taken into account in Section
6.2 of \cite{DEF1} and Eq.~(\ref{eq5.4}) above). Of particular
interest would be: (i)~the opening of a new, significant energy-loss
channel in (spherically symmetric!) gravitational collapse and
neutron-star binary coalescences; (ii)~important modifications in the
conditions for the onset of radiative instabilities in fast-rotating
neutron stars (Chandrasekhar-Friedman-Schutz instability) \cite{CFS}.
Both issues are particularly worth of further study.

\acknowledgments
We thank J.H.~Taylor and S.E.~Thorsett for informative discussions.
The work of T.~Damour at the Institute for Advanced Study was
supported by the Monell Foundation. The work of G.~Esposito-Far\`ese
at Brandeis University was supported by Centre National de la
Recherche Scientifique; partial support by NSF grant PHY-9315811
is also gratefully acknowledged.


\begin{table}
\caption{Critical baryonic mass $\overline m_{\rm cr}$
(and critical Einstein mass $m_{\rm cr}$) for some values of the
curvature parameter $\beta$ within the quadratic models
$A(\varphi) = \exp({1\over 2}\beta \varphi^2)$.}
\label{tab1}
\begin{tabular}{ccc}
$\beta$&$\overline m_{\rm cr}/m_\odot$&$m_{\rm cr}/m_\odot$\\
\tableline
           -10\hphantom{.34} & 0.69 & 0.66\\
\hphantom{1}-9\hphantom{.34} & 0.78 & 0.74\\
\hphantom{1}-8\hphantom{.34} & 0.89 & 0.84\\
\hphantom{1}-7\hphantom{.34} & 1.04 & 0.98\\
\hphantom{1}-6\hphantom{.34} & 1.24 & 1.16\\
\hphantom{1}-5.5\hphantom{4} & 1.38 & 1.28\\
\hphantom{1}-5\hphantom{.34} & 1.56 & 1.43\\
\hphantom{1}-4.5\hphantom{4} & 1.84 & 1.65\\
\hphantom{1}-4.35            & 2.01 & 1.78
\end{tabular}
\end{table}

\end{document}